# Dynamic Modeling and Real-time Management of a System of EV Fast-charging Stations

**Dingtong Yang** [a]
E-mail: dingtony@uci.edu
ORCiD: 0000-0001-7377-4531

**Navjyoth J.S. Sarma** [a]
E-mail: nsarma.js@uci.edu

**Michael F. Hyland** [a], **Corresponding Author**
Email: hylandm@uci.edu
Phone: (949) 824-5084
ORCiD: 0000-0001-8394-8064
Institute of Transportation Studies, 4000 Anteater Instruction and Research Bldg. (AIRB)
Irvine, CA 92697-3600

**R. Jayakrishnan** [a]
E-mail: rjayakri@uci.edu

[a] University of California-Irvine, Civil and Environmental Engineering
   University of California-Irvine, Institute of Transportation Studies

December 16, 2020

# Abstract

Demand for electric vehicles (EVs), and thus EV charging, has steadily increased over the last decade. Studies suggest that fast-charging facilities are crucial for the EV market. However, there is limited fast-charging infrastructure in most parts of the world to support EV travel, especially long-distance trips. The goal of this study is to develop a stochastic dynamic simulation modeling framework of a regional system of EV fast-charging stations for real-time management and strategic planning (i.e., capacity allocation) purposes. To model EV user behavior, specifically fast-charging station choices, the framework incorporates a multinomial logit station choice model that considers station charging prices, expected wait times, and detour distances. To capture the dynamics of supply and demand at each EV fast-charging station, the framework incorporates a multi-server queueing model in the simulation. The study assumes that multiple fast-charging stations are managed by a single entity (public or private) and that the demand for these stations are interrelated (through the station choice model). To manage the system of stations, the study proposes and tests dynamic demand-responsive price adjustment (DDRPA) schemes based on station queue lengths. The study applies the modeling framework to a system of EV fast-charging stations in Southern California. The computational results indicate that DDRPA strategies are an effective mechanism to balance charging demand across fast-charging stations. Specifically, compared to the no DDRPA scheme case, the quadratic DDRPA scheme reduces average wait time by 26%, increases charging station revenue (and user costs) by 5.8%, while, most importantly, increasing social welfare by 2.7% in the base scenario. Moreover, the study also illustrates that the modeling framework can evaluate the allocation of EV fast-charging station capacity, to identify stations that require additional chargers and areas that would benefit from additional fast-charging stations.

# 1 Introduction

## 1.1 Motivation

The usage of electric vehicles (EVs) has steadily increased over the past decade (Burnham et al., 2017) with the total number of EVs on the road in the United States (US) surpassing 1.3 million as of October, 2019 (Edison Electric Institute, 2019). Despite the recent growth in the EV market, EVs still represent a small proportion of light-duty vehicles in the US and across the world, with conventional gasoline-powered vehicles (GVs) representing the largest proportion. Given EVs (i) produce zero tailpipe emissions while GVs produce a variety and high-volume of harmful local pollutant emissions and (ii) EVs can be powered by renewable energy sources with minimal greenhouse gas emissions while GVs can only utilize greenhouse gas emitting fuel, the potential social and environmental benefits of converting the light-duty vehicle fleet from predominantly GVs to predominantly EVs are significant.

One of the main factors limiting the growth of EVs is the limited public and commercial EV charging infrastructure throughout the US and the world. A specific shortcoming of existing EV charging infrastructure is the lack of 'fast' chargers, also known as Level 3 chargers. CHAdeMO (2020) suggests a strong positive correlation between the number of fast chargers installed in cities in Japan and the number of EVs sold in those cities, indicating an important connection between EV growth and fast-charging infrastructure. Level 3 chargers are 8 to 12 times more efficient than Level 2 'slow' chargers. While Level 1 and Level 2 chargers are often used for at-home or at-destination (e.g., the workplace or shopping center) charging, Level 3 fast-charging enables en-route (or mid-trip) charging. Moreover, Level 3 charging is preferable for quick charging, long distance travel, and emergency charging, where charging time is a critical concern (Jabeen *et al.*, 2013). Fast-charging services are also beneficial for households living in high-density residential buildings where home charging may not be available.

The dearth of fast-charging facilities in the US is compounded by the shorter driving distances and longer recharging times of EVs compared to GVs (see Section 2.1 for a full comparison of EVs and GVs). In fact, 'range anxiety' from shorter driving distances between charges as well as the (in)ability to recharge batteries during travel are major obstacles for EVs in the consumer market (Egbue and Long, 2012).

In 2019, the total number of EV charging stations in the US reached around 28,400 with nearly 94,000 outlets (U.S. Department of Energy, 2020). The state of California, which is the largest EV market in the US, has more than 6,900 stations and more than 29,500 outlets (U.S. Department of Energy, 2020). However, only 14.6% of all charging stations in the US provide Level 3 fast-charging service and the current number of Level 3 fast-charging stations cannot fulfill the demand for fast charging throughout most of the country (U.S. Department of Energy, 2020). Potential problems associated with this supply-demand imbalance include long waiting times at fast-charging stations and/or long detour distances to find an EV fast-charging station.

Clearly, there is a significant need to invest in and expand EV fast-charging infrastructure capacity in order to support the growth of the EV market. This need is well known by both governmental agencies looking to support EVs and car manufacturers who are selling EVs. In addition to strategic planning, there is also a need to manage the EV fast-charging capacity efficiently at individual stations and across a region of fast-charging stations. Given the expected rapid growth rate of EVs over the next decade or two, there are likely to be imbalances between the supply of fast-chargers and the demand for fast charging, even as public sector and private companies work to expand fast-charging capacity. Managing supply and demand at the individual station level and across a regional system of fast-charging stations is important for minimizing user wait times at fast-charging stations and utilizing existing capacity effectively at fast-charging stations. Moreover, when demand outpaces supply in an area, demand-responsive pricing-based management strategies can generate revenue to support the funding and financing of additional fast-charging capacity to balance supply with demand in the long run.

## 1.2 Research Goal and Scientific Contributions

The goal of this research is to develop a dynamic modeling framework of EV fast-charging spatiotemporal demand and supply in order to (i) support the real-time management of a system of fast-charging stations and (ii) evaluate—in terms of service quality, charging system revenue, and social welfare—alternative systems of fast-charging stations under various demand settings for strategic planning support. While the long-term strategic planning of EV charging facilities through deterministic optimization modeling, often formulated as a siting and sizing problem (i.e., where should charging facilities be located and how many chargers should each facility have?), has received considerable attention in the literature (He *et al.*, 2013; Nie and Ghamami, 2013; Dong, Liu and Lin, 2014; Chung and Kwon, 2015; He, Yin and Zhou, 2015; Shahraki *et al.*, 2015; Ghamami, Zockaie and Nie, 2016; Chen, Liu and Yin, 2017; Huang and Kockelman, 2020; Sun, Chen and Yin, 2020), the real-time management of EV charging facilities through stochastic dynamic modeling of supply and demand has considerably less attention. To fill this gap, this study proposes an agent-based stochastic dynamic simulation modeling framework of a system of EV fast-charging stations. In addition to being dynamic, agent-based, and capturing stochasticity, the simulation modeling framework incorporates the following features:

- Each fast-charging station is modeled as a multi-server queue to capture the dynamics and stochasticity of demand and supply at individual stations.
- The manager agent of the system of fast-charging stations can dynamically adjust the prices at each station to reduce queue lengths through impacting EV user station choice behavior.
- Potential regional demand for fast-charging is exogenous (i.e. there is an upper bound) but demand for individual EV stations is endogenous.
- A discrete choice model captures a joint balking (i.e., choosing to not enter a queue/station and leave the system without charging) and station choice decision for EV user agents.
- EV user choices depend on the expected station waiting times, station prices, and station detour distances (i.e., the distance difference between an EV user driving directly between her trip origin and destination vs. detouring to a particular fast-charging station). The waiting times and station prices depend on (i) the actions of the manager agent, (ii) the choices of the other EV user agents, and the impact of i and ii on (iii) the current state of each station.

This study makes several notable contributions to the academic literature. As far as the authors are aware, this is the first study to model the dynamics of a regional system of multiple EV fast-charging stations that incorporates an EV user station choice behavioral model. The proposed model is a high-fidelity and high-resolution simulation model that is also computationally efficient. The major benefit of employing simulation methods is that they permit the analysis of complex interactions between EV user station choice decisions, the system manager pricing decisions, and service time stochasticity at fast-charging stations. Analytical models, while elegant and powerful, often fail to capture the complexities of real-world systems to the same extent as simulation models. Moreover, agent-based simulation models in general, and particularly the one developed in this study, can provide a wide-variety of output statistics for detailed analysis and decision support purposes (see Section 5 and 6).

A second major contribution is the modeling framework's integration of fast-charging supply and demand through multi-server queueing models of supply and a behavioral model of EV station choice (and balking). This integration should provide significant value to stakeholders tasked with evaluating potential long-term capacity expansion plans and/or the real-time management of a system of EV fast-charging stations, because, in the real-world, capacity allocation and pricing decisions directly and indirectly impact the behavior of potential EV customers.

A third major contribution is the incorporation of dynamic demand-responsive price adjustment (DDRPA) schemes in the modeling framework to manage the time-varying supply-demand imbalances at individual EV fast-charging stations and (indirectly) the whole system of EV fast-charging stations. The results illustrate the benefits and power of DDRPA schemes to manage spatial and temporal imbalances between supply and demand, in terms of EV user utility, charging system revenue, and overall social welfare.

## 1.3  Paper Outline

The rest of the paper is organized as follows. Section 2 provides background information on EVs and EV charging technology before reviewing the literature relevant to modeling and pricing EV charging infrastructure. Section 3 presents the proposed agent-based stochastic dynamic simulation modeling framework of a regional system of EV fast-charging stations. Section 4 presents the relevant input data and model parameters for the Southern California case study developed to validate the modeling framework and test the DDRPA schemes. Section 5 presents and discusses the results of case study analysis. Section 6 includes a scenario analysis to test the applicability of modeling framework and pricing strategies under different demand levels. The final section summarizes the paper, discusses the study's main implications, and presents future research directions.

## 2  Background and Literature Review

## 2.1  Electric Vehicles vs. Gasoline Vehicles

GVs dominate the personal vehicle market and are a relevant reference point to compare with the service quality offered by EVs. There are several key differences between EVs and GVs in terms of recharging and refueling that have implications for the planning and management of EV fast-charging infrastructure.

First, in one important way, recharging an EV is more flexible than refueling a GV. Unlike GVs that are almost exclusively refueled at dedicated gasoline refueling stations, an EV can be recharged in at least three different types of places—at a dedicated[1] EV recharging station, in a general parking lot with installed chargers (e.g., at workplaces, shopping centers, etc.), or at the EV owner's house. Home-installed chargers are typically Level 1 or Level 2 chargers while Level 3 chargers are only available at dedicated charging stations because of power supply and safety concerns. Home- and workplace-installed 'slow' chargers are mainly used for destination charging, while public fast-chargers aim to facilitate en-route charging, particularly for long-distance trips.

---

[1] 'dedicated' here refers to facilities whose main purpose is to recharge vehicles as compared with EV charging spots in general purpose parking lots where vehicles maybe recharge while the EV user conducts other activities (e.g., working, shopping, eating, etc.)

Second, the average range of an EV when fully charged is typically shorter than that of a similar GV. Although the average travel range of EVs has steadily increased in recent years, range-anxiety is still a concern for current and potential EV owners. The average range for an EV sedan is around 200 miles, which is only one-half to two-thirds the range of a typical GV sedan. This range issue has implications in terms of the frequency with which EV owners need to recharge and also the spatial distribution of EV fast-charging stations needed to support (long-distance) EV travel. Properly sizing and siting fast-charging stations is mainly a planning level challenge; however, these stations also need to be managed efficiently, particularly when demand outpaces supply (in certain areas, during certain times of the day).

Third, and most importantly from a real-time management perspective, the recharging time of an EV is considerably longer than the refueling time of a GV. Even with a 50kW Level 3 fast charger, to recharge an EV from empty to 80% capacity takes thirty minutes to an hour. Hence, if an EV does require the use of a EV fast-charging facility, the vehicle may occupy a charger at the station for a significantly longer period of time than a GV does at a gasoline station. Moreover, if an EV user goes to a recharging station with a queue, he/she may have to wait an extended period of time to even begin recharging. When a limited number of EV fast-charging stations are constructed along a high-volume corridor, a spike in local demand may create queues at one or more EV recharging stations, thereby inconveniencing EV users who need to recharge but do not have many alternative fast-charging stations nearby.

## 2.2   Electric Vehicles Type

There are two main types of EVs--Battery EVs (BEVs) and Plug-in Hybrid EVs (PHEVs). PHEVs include a second power source, such as gasoline, which means that PHEVs can refuel at a gas station or recharge at a charging station. Conversely, BEVs only have one power source—the battery. Since BEVs require re-charging and are the main users of public charging stations (the focus of this paper) the acronym EV refers to battery electric vehicles in this paper.

## 2.3   EV Charging Technology

The types of chargers usually found in public EV charging stations in North America are the J1772, the CCS (Combined Charging System), the CHAdeMO (abbreviation of "Charge de Move"), and the Tesla Supercharger. The first one is a Level 2 charger, while the other three include Level 3 Direct Current Fast Charging (DCFC) chargers (they are also capable of lower level charging). Table 1 summarizes EV charger types from Level 1 to Level 4. As mentioned previously, Level 3 chargers are 8 to 12 times more efficient than Level 2 chargers. Level 4 charging—also called Extreme Fast Charging (XFC)—is not currently available. Level 4 chargers could further improve charging efficiency and narrow the current refueling/recharging time difference between EVs and GVs.

Table 1 shows that compared with Level 2 chargers, Level 3 and 4 chargers have considerably more output power and therefore require higher voltage. The large loads from fast-charging stations will ultimately become a burden for electric grids and may cause system failures in the worst-case scenario. Studies have been done on both supply and demand sides of EV charging to avoid the situation. On the supply side, researchers propose methods of upgrading infrastructure and integrating fast-charging stations into the grid (Aggeler et al., 2010; Dharmakeerthi, Mithulananthan, & Saha, 2012; Guo, Deride, & Fan, 2016; Burnham et al., 2017; Z. Chen, Liu, & Yin, 2017; Iyer, Gulur, Gohil, & Bhattacharya, 2018). On the demand side, researchers use multiple approaches to model demand and propose control schemes (e.g. pricing) to balance the spatial and temporal demand for charging (Cao et al., 2012; Chen & Frank, 2001; Flath et al., 2014).

**Table 1 Summary of Common Level 1 to Level 4 EV Charging Facilities in North America**

| Level | Connector Type | Current Type | Voltage (V) | Power (kW) | Charging Time (Empty Battery) | Avg. Dist. per Min Charging (Mile/Min) |
|---|---|---|---|---|---|---|
| 1 | J1772 | AC | 120 | 1.4 | 250 miles/400km: 43 Hours | 0.1 |
| 2 | J1772 | AC | 240 | 3~20, typically 6 | 250 miles/400km: 11 Hours | 0.4 |
| 3 | CHAdeMO CCS | DC | Up to 500 | usually 50 | 80% of 250 miles/400km: 60 mins | 3.2 |
| 3 | Tesla Supercharger | DC | 400 | 120 | 80% of 315miles/500km: 40 mins | 5.6 |
| 4 | CHAdeMO 2.0 Revised CCS | DC | 800~1000 | > 400 | 250 miles/400km: 20 mins | 22 |

Authors summary and interpretation of multiple sources:
1. https://chargehub.com/en/electric-car-charging-guide.html#chargingconnectors
2. https://greentransportation.info/ev-charging/range-confidence/chap8-tech/ev-dc-fast-charging-standards-chademo-ccs-sae-combo-tesla-supercharger-etc.html
3. https://www.chademo.com/portfolios/sumitomo-electric-1-2/

## 2.4 Management and Pricing of EV Fast-charging Stations

This section reviews literature associated with managing EV charging stations via using different pricing strategies. One important application of pricing schemes is to manage and balance EV charging demand. These management schemes include Price Based (PB) and Demand Response (DR) Programs in electricity markets (Albadi and El-Saadany, 2008).

PB management programs are usually based on Time-of-Use (TOU) pricing or real-time pricing (RTP). TOU pricing involves charging users separate prices for peak and off-peak hours in order to shift loads from peak to non-peak periods. TOU prices are usually provided to users in advance; therefore, they cannot be used to respond to within-period demand stochasticity. TOU related studies include Cao *et al.*, (2012) and Bayram *et al.* (2015).

RTP involves dynamic price changes based on current demand levels. Chen & Frank (2001) study general service price adjustments based on the state of queues. Their study assumes that firms are capable of changing prices and that customers adjust behaviors based on queue states. Their results indicate that firms can increase surpluses while social welfare increases, under homogeneous customer behavior. Chen *et al.* (2017) design an RTP mechanism based on an automatic DR strategy for Photovoltaic-assisted charging stations. Luo, Huang and Gupta (2018) consider a case for stochastic dynamic pricing where renewable resources and energy storage are integrated to charging suppliers.

The management approach in this paper involves RTP under which fast-charging station prices are dynamically adjusted based on current demand levels; i.e. DDRPA schemes. It extends the setting in Chen

& Frank (2001) to a larger area with multiple EV fast-charging stations managed by the same entity. This paper describes schemes that enable heterogeneous pricing across stations based on the current state of each individual station, which is unlike the spatial or area pricing schemes used in earlier literature.

It is worth noting that several studies in the literature model the dynamic interplay between the electric grid power supply and EV charging station energy acquisition for management purposes (Dharmakeerthi, Mithulananthan and Saha, 2012, 2014; Sbordone *et al.*, 2015; Khan, Ahmad and Alam, 2019). Future research may combine the user-and-EV station dynamic models and management strategies in the current paper, with the EV station-and-power grid dynamic models and management strategies in the literature.

## 2.5   EV Charging Demand and Station Choice Behavior

Understanding and effectively capturing spatial-temporal demand patterns and EV user behavior is a crucial component for developing models to support the real-time management of a system of EV fast-charging stations. The relevant behavioral attributes in much of the EV station choice literature include, charging cost (Jabeen et al., 2013; Wen, MacKenzie and Keith, 2016; Daina, Sivakumar and Polak, 2017; Ge, MacKenzie and Keith, 2018), distance or detour distance to charging stations (Daina, Polak and Sivakumar, 2015; Sun, Yamamoto and Morikawa, 2016; Yang et al., 2016), as well as EV state of charge (SOC) and/or charging time (Sun, Yamamoto and Morikawa, 2015; Yang et al., 2016; Xu et al., 2017; Pan, Yao and MacKenzie, 2019). The model in the current study considers the following factors: charging price at each EV fast-charging station, detour distance to visit each station, and expected wait time at each station.

The behavioral modeling approaches in the literature include the multinomial logit (MNL) (Jabeen et al., 2013; Daina et al., 2017), the mixed logit (Sun, Yamamoto and Morikawa, 2015, 2016; Xu et al., 2017) and the nested logit (Yang et al., 2016), which are all discrete choice modeling approaches. Other modeling approaches for charging behavior include those by Kang & Recker (2009) who use an activity-based model and by Hu, Dong, & Lin (2019) who model choices based on cumulative prospect theory. In Jabeen *et al.* (2013), the charging choice is whether to charge at home, at work, or at public stations. The three alternatives are independent, as such the study employs the MNL model. The mixed logit model is appropriate when there is unobserved heterogeneity among users, differences in tastes, and/or when using panel data (Sun, Yamamoto and Morikawa, 2015). In Yang *et al.* (2016), the choice of charging is integrated with routing and the authors employ a nested logit model for route and charging choices. See Train (2009) for an overview of discrete choice models and their applications.

The current study employs the MNL choice model as it most appropriately matches the setting in the agent-based stochastic dynamic simulation. The simulation inherently assumes that a stochastic rather than deterministic model of station choice is appropriate, as traveler behavior in the real-world is likely to be uncertain. Moreover, because the modeling framework assumes the travelers are homogenous in their attributes and there are no additional attributes associated with EV charging stations aside from the ones considered in the model (price, queue length, and driving distance), the underlying assumptions of the MNL hold. Future research that has access to data associated with EV traveler behavior and the attributes of EV charging stations in a region may need to estimate a nested or mixed logit model and then incorporate these more complex choice models into the proposed framework in this paper. However, such data does not currently exist for public or (likely) private usage. Hence, the MNL model is an appropriate model to capture stochastic EV user station choice.

## 2.6 Queueing Models

To capture the stochasticity and dynamics of supply and demand over time at individual fast-charging stations, this study employs queueing models (Gross et al., 2008), which are quite common in the academic literature for EV charging studies (Bae and Kwasinski, 2012; Said, Cherkaoui and Khoukhi, 2013; Liang et al., 2014; Bayram et al., 2015; Farkas and Prikler, 2017). The $M/M/c$ queueing process is the most commonly used model in literature (Bae & Kwasinski 2012 and Farkas & Prikler 2012). Bayram et al. (2015) assume an $M/M/c/k$ queueing system (limited capacity of waiting room) and consider the probability of not entering the queue. In another study, Said et al. (2013), using a network queueing model approach, model every charging station as an $M/M/c$ queue and combine charging stations parallelly to construct a charging network. Liang et al. (2014) model charging as a three-step process (deciding, routing, and charging) and apply a BCMP queueing network model to estimate demand at each step.

The $M/M/c$ queue provides a closed analytical form for system analysis (queue length, total time in the system, etc.). By assuming a Poisson arrival process and a negative service rate, an $M/M/c$ model provides a closed form equation for queue length. However, the constant arrival rate assumption may not hold for a real-world system of EV charging stations. The more crowded an EV fast-charging station is and the longer the waiting time is, the lower the probability that it will be chosen by EV owners. In such a scenario, an EV user may drive to an alternative station or abandon the EV fast-charging system altogether and use Level 1 or 2 charger or simply not recharge at the current time. Therefore, the arrival rate at an individual EV fast-charging stations depends on the state of the charging station as well as the state of other charging stations in the region. Previous studies explore state-dependent service queues where the service rate depends on queue length (Takine & Hasegawa, 1994; Jain & Smith, 1997). Without the Poisson arrival process, obtaining an analytical form for arrival rate, queue length and waiting time is hard.

Instead of assuming an arrival rate for individual fast-charging stations, this study assumes a spatially- and temporally-varying Poisson arrival process for EV trips with two spatial attributes – the origin and destination locations of EV user trips. The study simulates the arrival process of EV user trips along with a set of EV user behavioral rules and EV system manager policies to endogenously determine the arrival rate and queueing dynamics at each fast-charging station in the system. It is worth noting that the EV system manger can charge/change prices to impact demand and thereby change the state of the system.

Chen and Frank (2001) analyze the queueing behavior of customers and their results indicate that for customers with homogenous behavior, for a single queue, adjusting prices based on state variables improves social welfare. The current study obtains similar results. However, the current study extends the one-station case in Chen and Frank (2001) to a system of EV fast-charging stations (i.e., a system of multi-server queues). The extension from a single queue to a system of multi-server queues required the implementation of a simulation model of the queueing system, rather than the analytical model of a single queue in Chen and Frank (2001). Finally, while Chen and Frank (2001) model balking using a single condition: 'is price charged currently larger than a fixed reserved price?', the current study adopts a MNL model to simultaneously model balking and charging station choice, considering multiple attributes including station price, expected station wait time, and detour distance to the station.

# 3 Modeling Approach

## 3.1 Overview

This paper proposes an integrated dynamic modeling framework for a regional system of EV fast-charging stations. Figure 1 depicts the overall modeling framework, while Table 2 summarizes the notation used throughout the paper. The proposed modeling framework integrates discrete choice models for EV user fast-charging station choice, queueing models to capture the dynamics of charging supply and demand at individual stations, and dynamic demand-responsive price-adjustment (DDRPA) schemes to manage each station. These components are integrated into an agent-based stochastic dynamic simulation model.

**Table 2 Notation Table**

| Notation | Description |
| --- | --- |
| $V$ | Electric vehicle set |
| $v_j$ | Individual electric *vehicle j* |
| $L_j$ | Current location of *vehicle j* |
| $D_j$ | Destination of *vehicle j* |
| $\mu_j$ | Charging time of a *vehicle j* |
| $\tilde{S}$ | Fast-charging station set |
| $S_i$ | Individual fast-charging station |
| $P_i$ | Price of a fast-charging *station i* |
| $P_{(i,t)}$ | Price of a fast-charging *station i* at time $t$ |
| $Det_i$ | Detour distance for charging when using *station i* |
| $W_i$ | Waiting time at a *station i* |
| $W_i^t$ | Waiting time at a *station i* at time $t$ |
| $\beta_1$ | Coefficient for price in MNL |
| $\beta_2$ | Coefficient for detour distance in MNL |
| $\beta_3$ | Coefficient for waiting time in MNL |
| $\lambda_i^t$ | Arrival rate of a fast-charging *station i* at time $t$ |
| $Q_{(i,t)}$ | Queue length of *station i* at time $t$ |
| $\alpha$ | Adjustment factor in price schemes |

The problem is defined in a general network $G = (N, A)$. An EV ($v_j$) is characterized by its current location node ($L_j$) and its destination node ($D_j$), where $L_j, D_j \in N$. Each EV is initiated with an initial state of charge (SOC) and a threshold for recharging. Fast-charging stations are located throughout the network at locations $\tilde{S} = \{S_1, S_2, S_3 \ldots S_i\}$, where $S_i \in N$. Each station $S_i$ has $c_i$ number of identical Level 3 chargers available.

The upper box in Figure 1 presents the EV user logic flow. When an EV agent needs to charge, it will first determine a set of feasible fast-charging stations based on the EV's current SOC and the EV user's willingness to detour. Given the charging station choice set, the EV agent chooses one of the stations (or to not use a fast charger) based on three parameters: charging station prices ($P_1, P_2, \ldots, P_i$), detour distances for users to access individual charging stations ($Det_1, Det_2, \ldots, Det_i$), and expected waiting times at stations ($W_1, W_2, \ldots, W_i$). The study employs the multinomial logit (MNL) model to capture charging station choice.

When an EV user arrives at a station, if the station has a vacancy, the EV user will be served immediately. Otherwise, the customer will be held in a single first-come, first-served queue at the station. Immediately prior to joining the queue, the EV user can evaluate the attributes of the originally chosen fast-charging station against the other charging stations in the system and determine if he/she wants to change stations, conditional on the EV's SOC allowing it to travel to another station. Once an EV user joins the queue at a specific charging station, he/she will not abandon the queue and the user locks-in the current charging price. EV users have a maximum number of times that they reroute—this parameter is set to one in this study. The EV user can also opt for 'no fast charging' (i.e., to leave the fast-charging system) if the 'no charging' alternative provides a lower disutility than any of the charging station options.

After joining a queue, the user must wait until the number of available chargers is greater than or equal to one. Although not shown, this wait time is calculated based on a simulation model of the multi-service queueing dynamics at the chosen station. After exiting the queue to access a server, the customer's service time, $\mu_i$, is determined based on two factors: the fixed time needed to acquire enough charge to reach the EV user's destination, and a random term to denote the extra charge a user may need/want to make subsequent trips after they reach their next destination.

The lower box in Figure 1 summarizes the charging station information flows. When an EV user joins the queue at their chosen charging station, the queue length increases by one unit at that station. Like any queueing model simulation, the one in this study tracks the agents in the queue and at the servers. Once a charging station has a queue with non-zero length, the DDRPA schemes are activated. The price and waiting time (calculated based on queue length) information for each charging station feed into the MNL station choice model that impacts the station choice of users who are making station choice decisions in the next time step.

Figure 1 omits the temporal dynamics of the system for clarity purposes. Nevertheless, the OD-based EV user trip demand rate is time-varying, and the simulation model tracks the state of each EV user and the state of each charging station (and charger) throughout the day. The state of each EV user is characterized by their location and their SOC. The model tracks EV user movements to stations, their time in the queue, and their time in-service. The state of each EV charging station is characterized by its price and its queue length. The EV system manager dynamically adjust prices at individual stations based on the state of each individual station.

The following three subsections provide more details on the modeling components in Figure 1. The next subsection details the EV user station choice model. The following subsection details the queueing model with state-dependent arrivals. The last section provides an overview of the three DDRPA schemes.

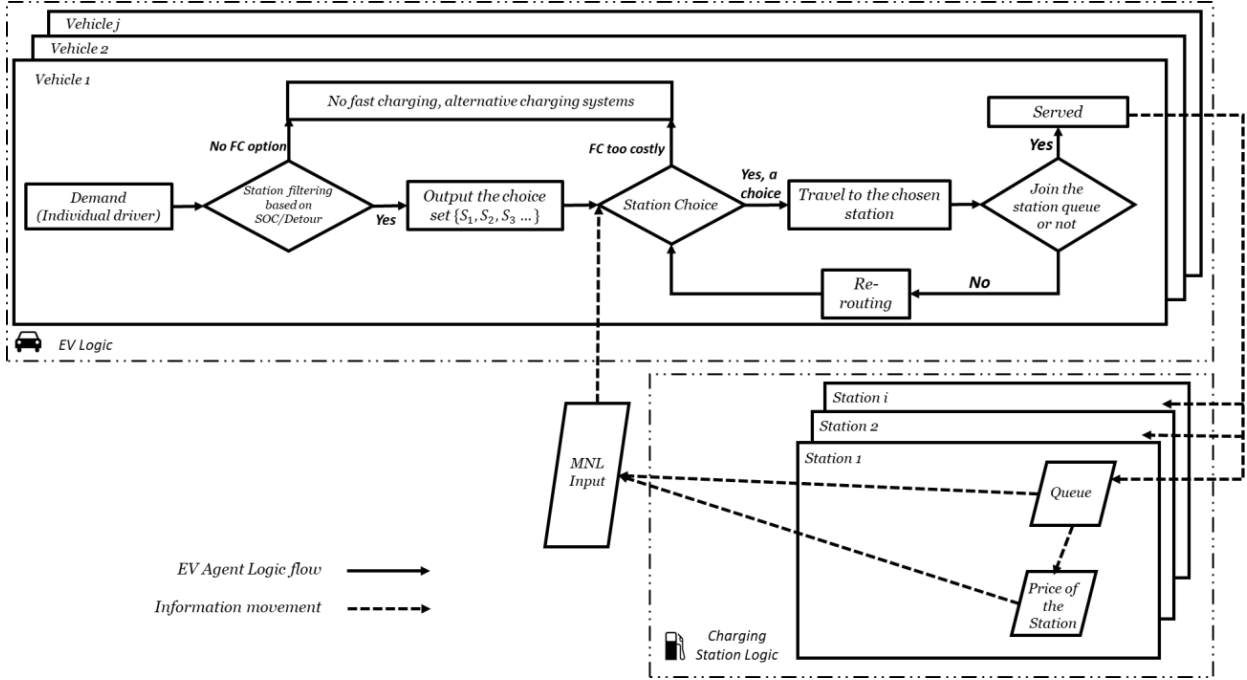

*Figure 1. Overall modeling framework*

## 3.2 EV Fast-charging Station Choice Models

This subsection describes the behavioral models of the EV user station choice that underlay the demand for fast-charging stations. As mentioned previously, the study employs an MNL model to capture the station choice (or 'no station' choice) of individual EV users. However, another behavioral model is required to determine each EV user's station choice set.

The behavioral model used to determine the EV user's choice set is a straightforward deterministic model with two parameters—the EV user's current SOC and his/her willingness-to-detour. Figure 2 displays how the deterministic model, with the SOC and willingness-to-detour parameters, produces a choice set for each EV user. The subregion delineated by the yellow circle includes the set of charging stations the EV can reach based on its current SOC ($S_1, S_2, S_3, S_4$). The subregion delineated by the blue ellipse includes the set of charging stations the EV is willing to visit between its current location and its destination given his/her willingness to detour ($S_2, S_3, S_4, S_6$). The intersection of the blue ellipse and the yellow circle includes all the charging stations in the EV user's choice set ($S_2, S_3, S_4$).

After determining a set of feasible charging stations, each EV user chooses an individual charging station $S_i$ considering the price, detour distance, and expected waiting time of each charging station in his/her choice set. This study assumes the initial station choice is made when the EV user's trip begins based on the current information about station prices and queue lengths, available through a mobile or desktop application. When an EV user arrives at a station, the study assumes he/she reconsiders his/her station choice before entering the queue. In all cases, station choices are based on the MNL model. A future model extension might assume EV users regularly re-consider their station choice while driving based on evolving information about queue lengths and charging prices at stations in the system.

Based on the MNL assumptions, Eqn. (1) displays the probability of choosing station $S_i$ as a function of the three variables ($P_i, Det_i\ and\ W_i$) and the parameter values ($\beta_1, \beta_2, \beta_3$) associated with each variable.

$$Pr(S_i) = \frac{e^{\beta_1 \times P_i + \beta_2 \times Det_i + \beta_3 \times W_i}}{\sum_{k=1}^{K} e^{\beta_1 \times P_k + \beta_2 \times Dis_k + \beta_3 \times W_k}} \quad (1),$$

where $P_i, Det_i\ and\ W_i$ represent the current charging price at Station $i$, total detour trip distance to use Station $i$, and current expected waiting time at Station $i$, respectively.

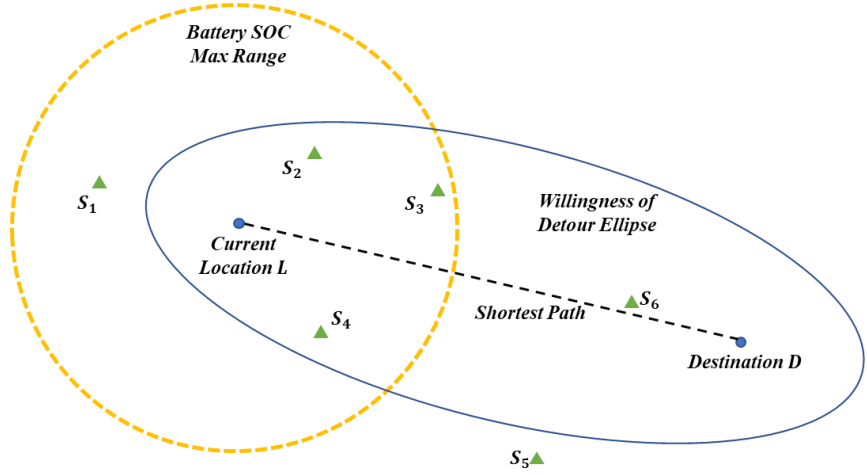

*Figure 2: Determining an EV User's Charging Station Choice Set*

## 3.3 A Queueing Perspective of the Modeling Framework

The modeling framework could also be viewed from the perspective of a system of charging station queues. The system metrics of queueing models (e.g., queue length and waiting time) are obtained for the evaluation of a system of EV fast-charging stations. In the case where multiple charging stations in an area are owned by a single entity, this study hypothesizes that the entity can increase revenues and balance charging demand across stations by employing a DDRPA scheme. Figure 3 depicts the basic dynamics of price and arrival rate at charging stations.

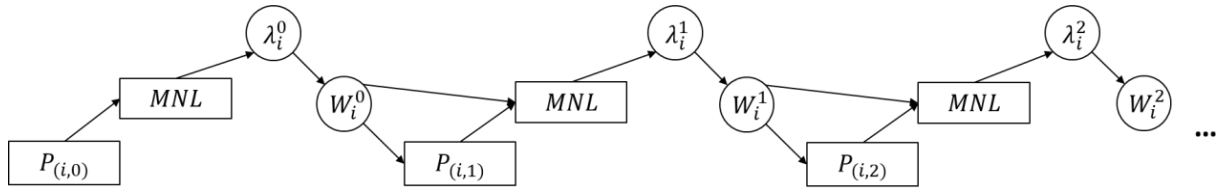

*Figure 3: Price and arrival rate changing dynamics of charging stations*

Previous research, described in the literature review, indicates that EV owners choose charging stations based on multiple factors including price and waiting time. For a charging station, when it has a higher price than others or a longer expected waiting time, the charging station is less likely to be chosen. The arrival rate function for a charging station $i$ at time $t$ could be written as:

$$\lambda_i^t = f\left(P_{(1,t)}, P_{(2,t)}, \ldots P_{(i,t)}, W_1^{t-1}, W_2^{t-1}, \ldots W_i^{t-1}, \text{Spatial Factor}\right) \quad (2)$$

The arrival rate of charging station $i$ is a function of charging price of station $i$ and the price of other stations in the system, the waiting time at station $i$ and the waiting time of other stations in the system, as well as the spatial distribution of the charging stations relative to the spatial-temporal distribution of origin-destination demand flows. Eqn. (2) implies a complex relationship between the arrival rate of a station and the factors on the right-hand side, which likely does not admit a closed analytical form. Figure 3 also shows that the arrival rate at station $i$ at time step $t$ is not directly related to the prices and waiting time at time $t-1$, but through the MNL choice model. Instead of assuming an arrival pattern of vehicles at individual stations, we simulate the arrival process as shown in Figure 1.

Under the modeling framework mentioned in Section 3.1, the charging station entity can employ pricing schemes to encourage/discourage the arrival of users at individual charging stations and therefore achieve the managerial goal of balancing area demand. A balanced charging system could potentially improve system performance. Customer waiting time and disutility could be reduced. On the contrary, if no pricing schemes are employed, the demand-supply balancing mechanism merely relies on customer's observation of queue lengths and the associated expected waiting time at individual stations, which may be an inefficient and ineffective balancing procedure.

In Figure 3, if a charging station $i$ has initial price $P_{(i,0)}$. The price $P_{(i,0)}$, together with the charging price of other stations will determine the arrival rate $\lambda_i^0$. The expected waiting time is determined by the arrival rate, jointly with the service rate, at the end of time period 0. At time period 1, the arrival rate is affected by both the current price $P_{(i,1)}$ and expected waiting time, which comes from the queue length from the previous time step. The two factors jointly determine the arrival rate at time step 1, $\lambda_i^1$, which subsequently impacts queue length at the end of time period 1. Section 5 displays the temporal dynamics of queues, prices, and waiting times at EV fast-charging stations for illustrative purposes.

In other applications, the price may also depend on the input cost of electricity from the electric grid; however, this is beyond the scope of the current study. The 'price' in this study should be thought of as a base fee for access to a charger plus a congestion charge.

Figure 4 presents a toy case. At time step 0, four vehicles $(v_1, v_2, v_3, v_4)$ request charging. After the MNL-based station choice model, $v_1\ and\ v_2$ choose Station $S_1$, $v_3$ chooses $S_3$, while $v_4$ decides not to charge at a fast-charging station and leaves the system. At time step 1, both $v_1\ and\ v_2$ arrive at $S_1$. Since $v_1$ arrives first and is served first, $v_2$ decides not to wait for $v_1$ to finish but heads to an alternative station, $S_3$. Meanwhile, new charging demand $v_5, v_6, v_7$ enter the system and choose their respective charging stations. Since $S_1$ is occupied and full, it becomes less favorable for the new EVs. At time step 2, vehicle $v_8\ and\ v_9$ request charging. The station operator decides to increase the price of $S_1$ (not shown in the figure) to discourage the new arrival at $S_1$. Originally the preferred charging location for $v_8$ was also $S_1$. However, with the increased price, $v_8$ decides to switch to $S_2$ to avoid potential excessive waiting.

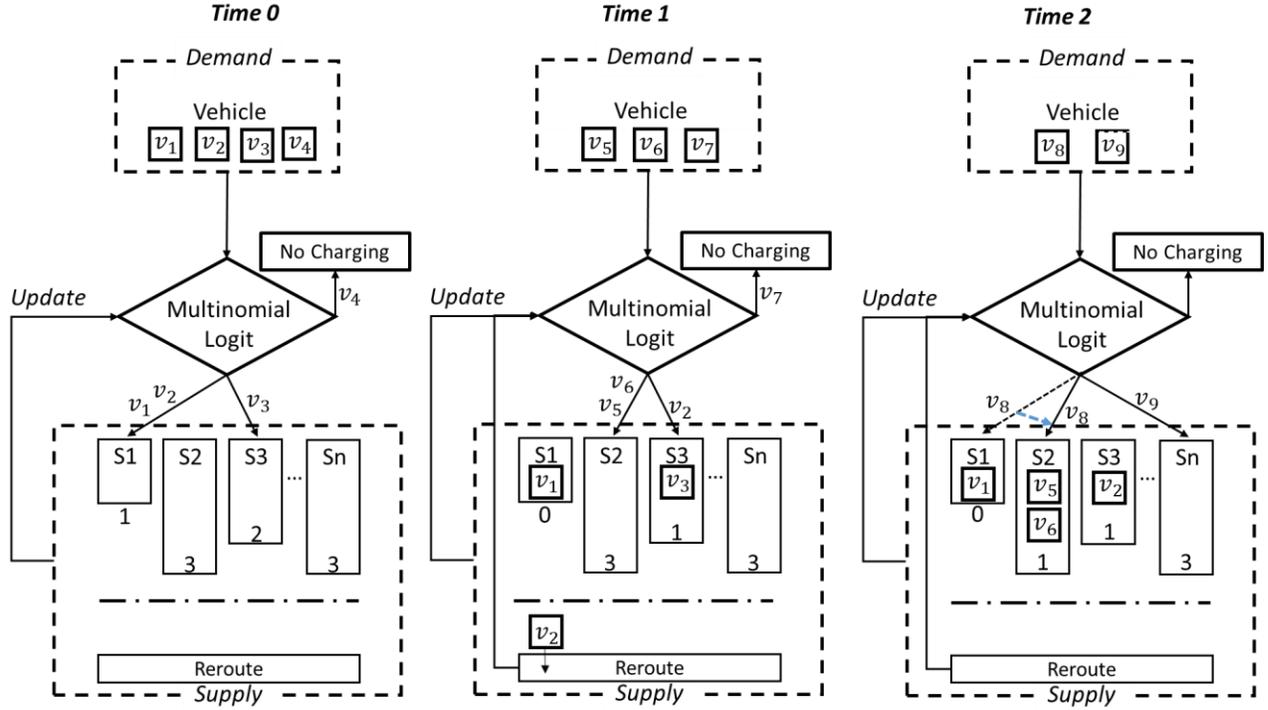

*Figure 4: Dynamics of the queueing system*

## 3.4 Queue-based Station Price Adjustment Schemes

The service queue at each EV charging station has a self-balancing mechanism. A higher arrival rate results in a longer queue, while a longer queue leads to a decrease in the arrival rate. However, EV fast-charging station operators can also balance supply and demand in the system through dynamically adjusting prices at each station. Using DDRPA schemes to manage supply and demand in the system rather than allowing the system to self-balance based on queue lengths should allow the operator to significantly increase revenues and decrease user wait times without significantly decreasing the utility/welfare of EV users.

For station $i$, the price adjustment schemes considered in this study include:

$$P_{(i,t)} = P_{(i,0)} + \alpha_1 \times Q_{(i,t)}, (Linear\ Adjustment); (3)$$

$$P_{(i,t)} = P_{(i,0)} + \alpha_2 \times Q_{(i,t)}^2, (Quadratic\ Adjustment); (4)$$

$$P_{(i,t)} = P_{(i,0)} + e^{\alpha_3 \times Q_{(i,t)}} - 1, (Exponential\ Adjustment); (5),$$

where $P_{(i,0)}$ is the base price of the station $i$. $Q_{(i,t)}$ is the number of customers waiting in the queue of station $i$ at time $t$. In practice, to avoid excessively frequent price change with respect to queue length, a charging station may impose a step function for adjustment, which could be written as:

$$P_{(i,t)} = P_{(i,0)} + \alpha_1 \times floor\left(\frac{Q_{(i,t)}}{m}\right), (Linear\ Adjustment); (6)$$

$$P_{(i,t)} = P_{(i,0)} + \alpha_2 \times floor\left(\frac{Q_{(i,t)}^2}{m}\right), (Quadratic\ Adjustment); (7)$$

$$P_{(i,t)} = P_{(i,0)} + e^{\alpha_3 \times floor(\frac{Q_{(i,t)}}{m})} - 1, (Exponential\ Adjustment); (8)$$

where the number $m$ is the step factor for updating prices, which can be chosen according to the service rate and the number of chargers at a specific station. In this study, price is updated when queue length increases, and the updated price is sent to the MNL station choice model for the next time period. Moreover, in the case study defined in Section 4, $m$ is set to 3, and prices are adjusted after every third arrival at a station.

## 4 Case Study

### 4.1 EVgo Regional EV Fast-charging System

To validate the modeling approach and demonstrate the effectiveness of the DDRPA schemes, this study models the EVgo system of fast-charging stations in the combined area of Los Angeles County and Orange County (Total: $5,699\ miles^2, 14,760\ km^2$) in California, US. The case study includes all 82 fast-charging stations (288 chargers) owned by EVgo in the study region (EVgo, 2020). EVgo claims to own the largest public fast-charging network in the US, and as such, their network is the best representation of a regional system of EV fast-charging stations managed by the same entity.

The number of chargers at each individual station was retrieved from the US Department of Energy Website in September, 2019 (U.S. Department of Energy, 2020). Figure 5 shows the entire study area with all fast-charging station locations marked by various colored dots. The dot colors (and sizes) denote the number of Level 3 DCFC chargers at each station. The number of chargers for a single station ranges from 1 to 8, and the average number of chargers per station is 3.5 ($\sigma = 1.28$).

### 4.2 Model Parameters and Data

The California State Travel Demand Model (CSTDM, model year 2020) is used to generate EV trips that may require fast-charging. Demand is generated for the AM peak (6 am to 10 am), Mid-day period (10 am – 3 pm) and PM peak period (3 pm – 7 pm). The trip rate varies significantly across the three periods, but the study assumes the trip rate is constant within each period. The study is based on trips that either originate, terminate, or pass-through Los Angeles and Orange County, California. All short and long-distance trips within the study area are obtained directly from CSTDM. In addition, a sub-area analysis was conducted using TransCAD for trips which pass-through the study area whose origins or destinations fall outside the boundaries of the study area.

Table 3 displays the relevant parameters in the modeling framework for the case study. As mentioned previously, the study assumes Battery Electric Vehicles (BEVs) to be the primary users of public charging stations. In 2019, roughly 1.4% of all registered automobiles in California were BEVs (California Department of Motor Vehicle, 2020). The base case scenario assumes 1.4% of all vehicles in California will use public charging stations (the base scenario). The 1.4% factor is applied uniformly to the CSTDM demand across all zones in the study area. In addition, Section 6 includes scenario analysis where the market penetration rate varies across scenarios to test the applicability of the modeling framework and pricing schemes. The scenario analysis compares the base scenario to a lower demand scenario and a higher demand scenario.

The average trip rate for an O-D pair is based on the long- and short-distance O-D tables from the CSTDM. Trip generation follows a Poisson process over the study period with a mean rate equal to the average trip generation rate in the O-D tables, for each zone, and time period. The coordinate locations of the trip origins and destinations are assumed to be distributed uniformly within the origin and destination zones.

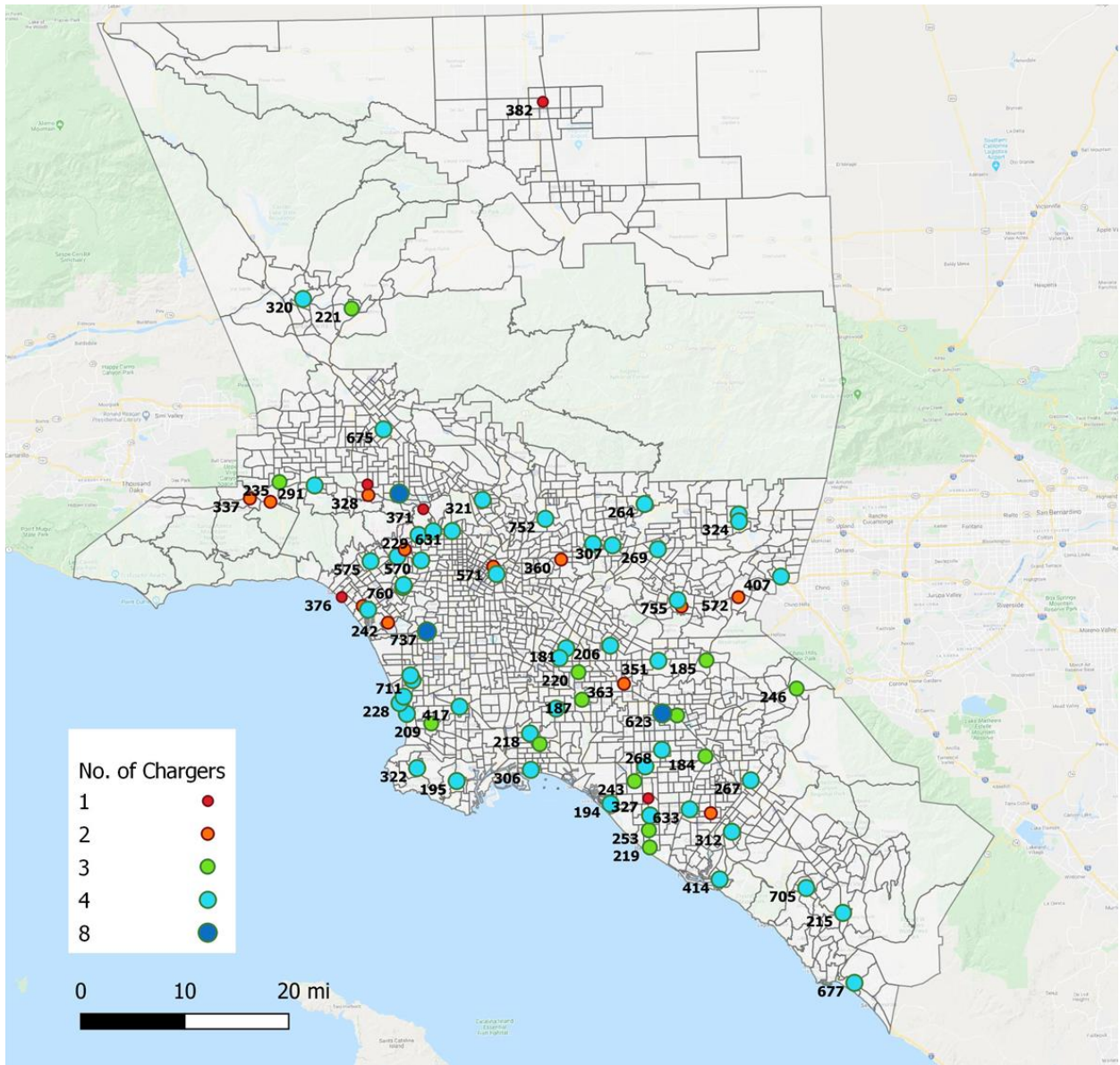

*Figure 5: Regional EV Charging System Case Study Area*

**Table 3: Summary of Simulation Parameters**

|  | Parameter | Value | Reference |
|---|---|---|---|
|  | Simulation Time | 6 a.m. to 7 p.m. |  |
| **MNL Model Parameters** | Coefficient for Price | -2.7 util. per dollar | Authors' interpretation: Lam and Small (2001), Yang et al (2016), Sun et al. (2016) and Motoaki and Shirk (2017). |
|  | Coefficient for Detour Distance | -3.2 util. per mile |  |
|  | Coefficient for Waiting Time | -1 util. per min |  |
|  | No Fast-charging Disutility | -50 util. | Authors' interpretation |
| **Supply** | Charging Stations | 82 stations | US DOE (2019) |
|  | Total No. of Chargers | 288 chargers | US DOE (2019) |
|  | Customer service time | Fixed + random | Motoaki and Shirk (2017) |
|  | Base Charging Rate* | $5 per hour | US DOE (2019) |
|  | Vehicle Speed | 50 mph |  |
| **Demand** | BEV Market Penetration | 1.4 % | California DMV (2020) |
|  | Total No. Charging Requests | 4,177 requests | CSTDM (2020) |
|  | Initial State of Charge (SOC) | $N(0.6, 0.2)$ | Sun et al (2016) |
|  | EV Recharging Threshold | $N(0.5, 0.1)$ | Hu et al (2018) |
|  | Maximum Detour Distance | 10 miles |  |

*The base rate was set as hourly based on DOE stats, the current charging plan for EVgo is based on minutes.

The initial state of charge (SOC) of EVs at the beginning of their trips follows a truncated normal distribution with $mean = 60\%, \sigma = 20\%$. Each vehicle has a recharging threshold – a level of charge below which the EV user wants to recharge. This recharging threshold follows a truncated normal distribution with $mean = 50\%, \sigma = 10\%$. The SOC status distribution is based on Sun, Yamamoto and Morikawa (2016) who show a real-world distribution of initial SOC before charging, with a mean between 50% to 60%. Sun, Yamamoto and Morikawa (2016) also indicate that people do not necessarily wait until a very low SOC to charge their vehicles and some people prefer to charge even at high SOC (e.g. 80%).

The service time parameter at EV charging stations contains two components. The first component is the necessary time for an EV user to charge in order to reach his/her next location/destination. This time is simply the minimum charge needed multiplied by the charging efficiency/rate. The second component is a random term that captures the uncertainty associated with how much additional charge the EV user wants to acquire. The random component follows a negative exponential distribution with mean = 10 mins. The charging time distribution has been compared with empirical evidence in Motoaki and Shirk (2017) to tune the parameter.

Choice sets of charging stations for each EV user are generated based on the behavioral model described in Section 3.1, which is a deterministic model that incorporates EV SOC and the user's willingness to detour (10 miles in this scenario). In total, the analysis includes 6,641 EV users with charging demand. The $demand/supply$ (of charging hours) ratio for the entire region is 0.37; however, the spatial distribution of demand and EV fast-charging stations throughout the region cause the ratio to vary substantially across the region. Distances are calculated by location coordinates using the Manhattan distance metric. The analysis assumes the average vehicle speed in the study area is 50 miles/hour.

The base rate for charging is $5 per hour according to DOE database. The simulation runs for 13 hours (6a.m. to 7 p.m.), which includes the AM Peak, midday, and PM Peak travel demands. The parameters in the station choice MNL model include $\beta_1 = -2.7, \beta_2 = -3.2, and \beta_3 = -1$ for price, detour distance, and wait time, respectively. The relative magnitudes of parameters in the MNL represent the relative weights of the attributes. Yang *et al.* (2016) find that when making charging and routing decisions, users value

travel distance (measured in kilometers) twice as much as time (measured in minutes). Negative one is assigned for the waiting time parameter and is used as a reference point. When using miles for the distance unit, the weight/parameter assigned to distance is $-3.2$. Based on Lam and Small (2001), who analyze the relative importance of price vs. time (i.e. value of time), one-hour travel time is equivalent to \$22, which indicates on average people value each dollar 2.7 as much as a minute of time.

A no-fast-charge option is also included in the choice model with a fixed disutility of -50. The reasoning for -50 is as follows. When choosing the no-charging option, a customer leaves the fast-charging system, and may use alternative charging options such as charging at Level 2 chargers. A 10-min fast-charging session will provide the same amount of electricity as a 60-min Level 2 charging session. Multiplying the 50-minute gap in charging time by the parameter for time lost (-1 util per min), results in a disutility of 50 units.

## 5  Results

This section presents the computational results for the case study described in Section 4. It is important to note that the input demand and the model parameters in Section 4 were not calibrated to match the existing EVgo system performance in Southern California. As such, the computational results in the current section only aim to illustrate the usefulness of the modeling framework for real-time management purposes and to evaluate the spatial allocation of fast-charging capacity given a reasonable spatiotemporal demand distribution. The results are not meant to directly inform the planning and management of the existing EVgo system of stations.

The first subsection analyzes the customers who choose to leave the EV fast-charging system without charging. The next six subsections compare the system metrics and analyze the impacts of the four DDRPA schemes on arrival rate, average queue lengths, station prices, total revenue, average customer wait times, and overall customer disutility before combining total revenues and overall customer disutility into a social welfare measure.

### 5.1  Analysis of Lost Customers

In a queueing system, unavoidably, some customers may leave the system without being served. The results in this analysis of fast-charging stations in Southern California indicate that whether a customer leaves the system heavily depends on the number of stations in their initial choice set. Hence, the total number of lost customers naturally depends on the spatial-temporal distribution of demand and spatial distribution of chargers. The formation of the choice set, as described in Section 3.2, depends on the EV's SOC and the user's willingness to detour, as well as the location of the EV user's trip origin and trip destination and the location of EV fast-charging stations.

Table 4 summarizes the number of lost customers and percentage of lost customers under different pricing schemes as a function of the size of the EV user's charging station choice set. The table shows that under the no price adjustment case 32.5% of customers with only one station in their choice set leave the system, whereas 8.1% and 0.4% of customers with 2 or 3+ stations in their choice set leave the system, respectively. The results are similar for the cases with price adjustment schemes; albeit the pricing schemes do slightly increase the number of lost customers. For EV users with only one charging station in their choice set, even without price adjustment schemes, around 30% are not served, which is an indication of poor spatial distribution of charging station capacity in the regional system.

**Table 4: Lost Customers and Percentages Results**

| Total Users: 4,177 | No Adjustment | | Linear | | Quadratic | | Exponential | |
|---|---|---|---|---|---|---|---|---|
| Number of Charging Station Choices | Lost Customer | % of the Category | Lost Customer | % of the Category | Lost Customer | % of the Category | Lost Customer | % of the Category |
| *1* | 430 | 32.5% | 464 | 34.7% | 476 | 35.4% | 477 | 35.7% |
| *2* | 74 | 8.1% | 80 | 8.9% | 100 | 11.4% | 91 | 10.0% |
| *≥3* | 7 | 0.4% | 3 | 0.2% | 7 | 0.4% | 6 | 0.3% |
| *Total* | 511 | | 547 | | 583 | | 574 | |

Figure 6 maps the origin-destination trip flows of served and lost customer, for the no pricing case. The left figure shows that the stations near the northern border (Area 1) and eastern border (Area 2) fail to accommodate the trips in those areas. In detail, there is only one charging station (Station 382) with one charger in Area 1. The customers in that area are lacking charging station options and the charging station lacks sufficient capacity to meet demand. For Area 2, though four stations exist, the number of chargers is not sufficient for the high demand in this area. When planning for new charging stations, the charging company may consider adding chargers at existing stations in Area 2 and/or building new charging stations around Area 1 and Area 2.

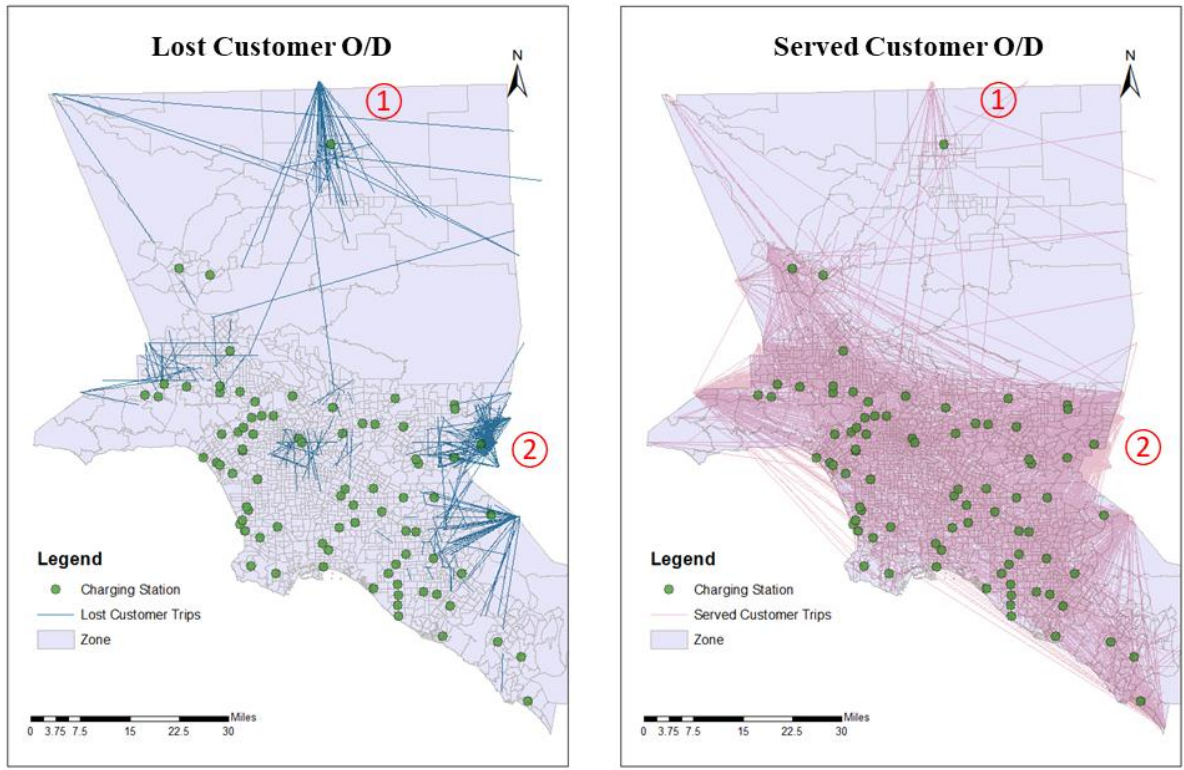

*Figure 6. Lost (left) and Served (right) Customer Origin-Destination Flows*

### 5.1.1 Sensitivity Analysis on Lost Customers with respect to the Demand:Supply Ratio

Besides comparing the number of lost customers under different pricing schemes, this section also tests the sensitivity of the percentage of lost customers with respect to changes in traffic intensity (i.e., the $demand/supply$ ratio). The $demand/supply$ ration is calculated by:

$$\frac{Demand}{Supply} = \frac{Number\ of\ EVs \times Average\ service\ time}{Total\ number\ of\ chargers \times Total\ time} \quad (9)$$

This metric provides an indication of the sufficiency of charging stations and chargers with respect to different demand levels. The analysis provides insights into whether lost demand is due to overall system capacity issues or simply the spatial imbalance between EV station chargers and trips flows.

Figure 7 displays the percentage of lost customers as a function of the $demand/supply$ ratio, where the ratio ranges from 0.1 to 1.0. The relative lost customer percentages (the y-axis on the right-side of Figure 7) are calculated as follows:

$$Relative\ \%\ of\ Lost\ Customers = \frac{No.of\ Lost\ Cust_{Price\ Adj} - No.of\ Lost\ Cust_{No\ Adj}}{Total\ No.of\ Customers} \times 100\% \quad (10)$$

Figure 7 shows that the percentage of lost customers and traffic intensity have an approximately linear relation. When the demand level is lower than 0.2, linear DDRPA reduces the number of lost customers. Results are similar for the exponential DDRPA. At low levels of demand, the quadratic DDRPA scheme produces a similar number of lost customers as the no pricing option. When the demand level is higher than 0.3, the linear DDRPA scheme results in a slightly higher lost customer percentage than the no-pricing case. The number of lost customers in the quadratic and exponential cases are between 1% and 3.5%.

Figure 7 also indicates that without price adjustments and at a medium level of $demand/supply$ ratio (e.g., 0.4), the regional charging system still loses around 15% of the total customers. This result is an indication of a spatial imbalance between EV fast-charging stations and the travel demand in the current system. The details of this imbalance are illustrated in Figure 6.

The findings presenting in this subsection illustrate that the proposed modeling framework can help identify (i) the stations that need more chargers, (ii) the users and the attributes of their trips that would benefit from increased access to EV fast-charging stations and chargers, and (iii) spatial regions that would benefit the most from new charging stations. The modeling framework can also help evaluate the impacts of new charging stations and/or additional chargers on all the system performance metrics.

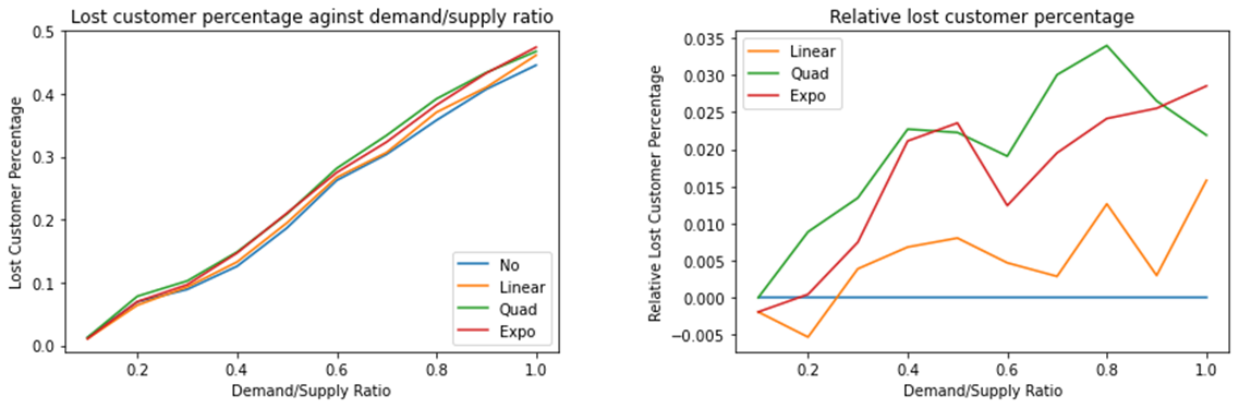

*Figure 7: Lost Customer Percentage vs Demand:Supply Ratio*

## 5.2 Station Queue Length Results

To evaluate the congestion level of a station and measure the performance of charging stations with different price adjustment schemes, this section compares queue lengths at the most congested station and across all stations. The analysis employs the same demand input file for each DDRPA scheme.

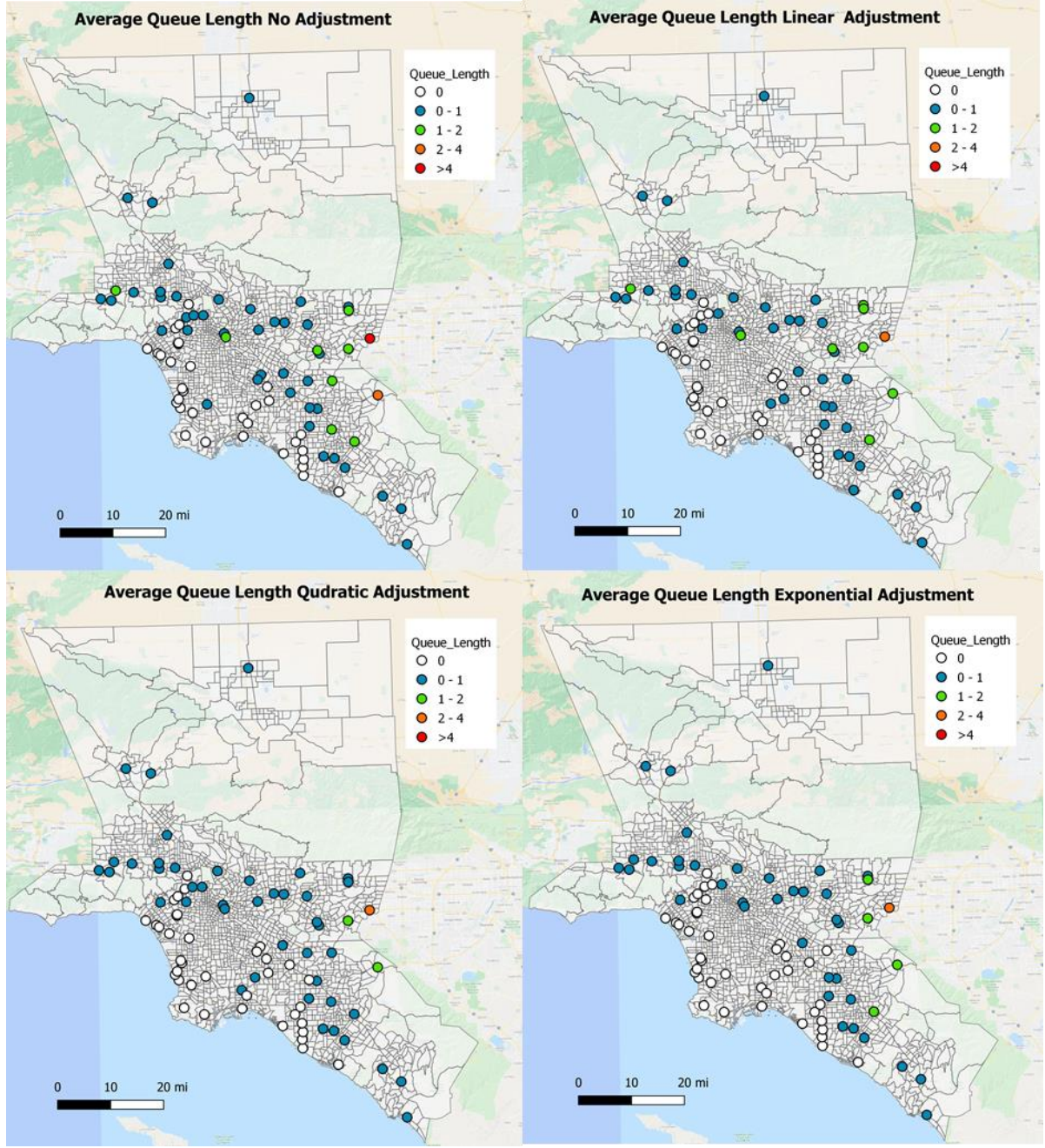

*Figure 8: Average Queue Lengths of Charging Stations across Price-adjustment Schemes*

Figure 8 displays the average queue length of each individual station under the four DDRPA schemes. While some stations have longer queues after imposing price adjustments, the overall system queue length decreases under DDRPA schemes compared to the no pricing case. The system-wide peak hour (including both AM and PM peaks) average queue lengths under the four cases are $0.40, 0.37, 0.29, 0.31$ person/ (station $\cdot$ minute) for the no-adjustment, linear, quadratic, and exponential schemes, respectively. Hence, the DDRPA schemes can, and do, significantly decrease queue lengths relative to the no price adjustment case. One potential explanation for the average queue reduction is the rebalancing mechanism of price adjustments. Interestingly, the number of zero-queue stations decrease when pricing schemes are implemented, indicating that the pricing schemes effectively shift demand from high utilization stations to low utilization stations. Of course, the DDRPA schemes may result in other negative system impacts related to users leaving the system and paying more for recharging; the next several subsections analyze these other system impacts.

As an illustrative example of how pricing impacts queue lengths, Figure 9 displays the queue length over time for a single station—Station 407. This is the most congested station in the entire regional system of charging stations. A clear pattern of peak and non-peak charging periods is evident. A peak charging period appears between 6:20am and 11:50am. Another peak charging period happens between 3:00pm. And 7:00pm. No intensive congestion is observed during the midday due to low overall demand. The queue length of Station 407 in the no pricing case fluctuates around 7 EVs during peak hours. The three DDRPA schemes decrease the queue lengths during peak hours, compared to the no pricing case. Comparing the four cases, both the quadratic and the exponential adjustments reduce the average queue length by approximately 36% compared to the no pricing case, while the linear adjustment reduces queue lengths by approximately 25%. Notably, there is always a queue in all four cases during peak hours indicating that the station operates at full capacity during the peak period, even with price adjustments. Hence, none of the price adjustment schemes seem to be overly responsive such that stations become under-utilized.

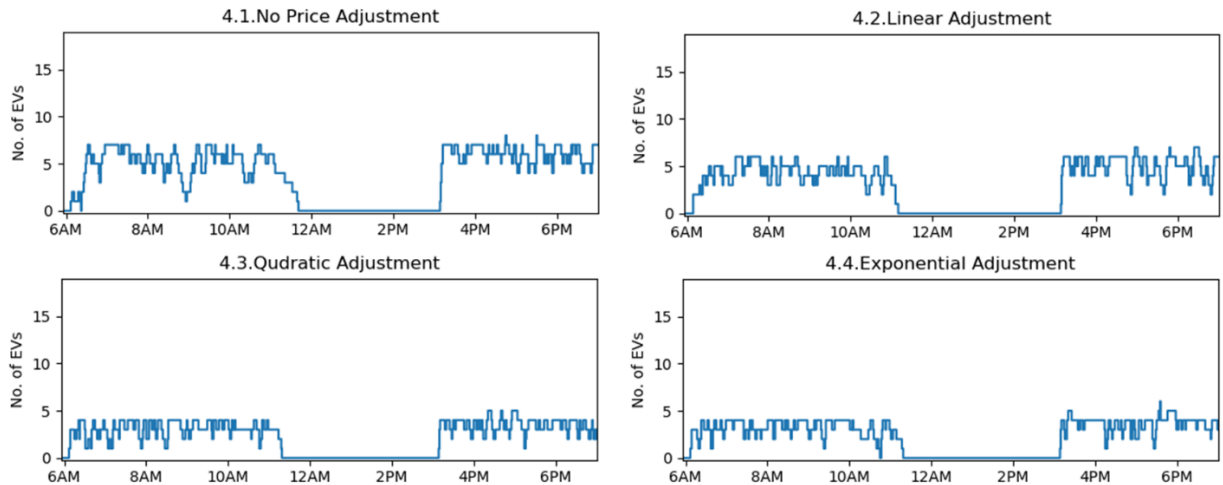

*Figure 9: Queue lengths under different price-adjustments schemes for Station 407*

## 5.3 Charging Prices

Figure 10 presents the time-varying arrival rate and prices for Station 235 to illustrate the relationship between arrival rate and station price. Comparing the first figure (no adjustment) with the other three, price adjustment schemes clearly reduce the arrival rate during the peak period at Station 235—a congested station. When the price increases, arrival rates decrease more dramatically. It is important to note that the arrival rate at Station 235 also depends on the prices and queue lengths at other stations.

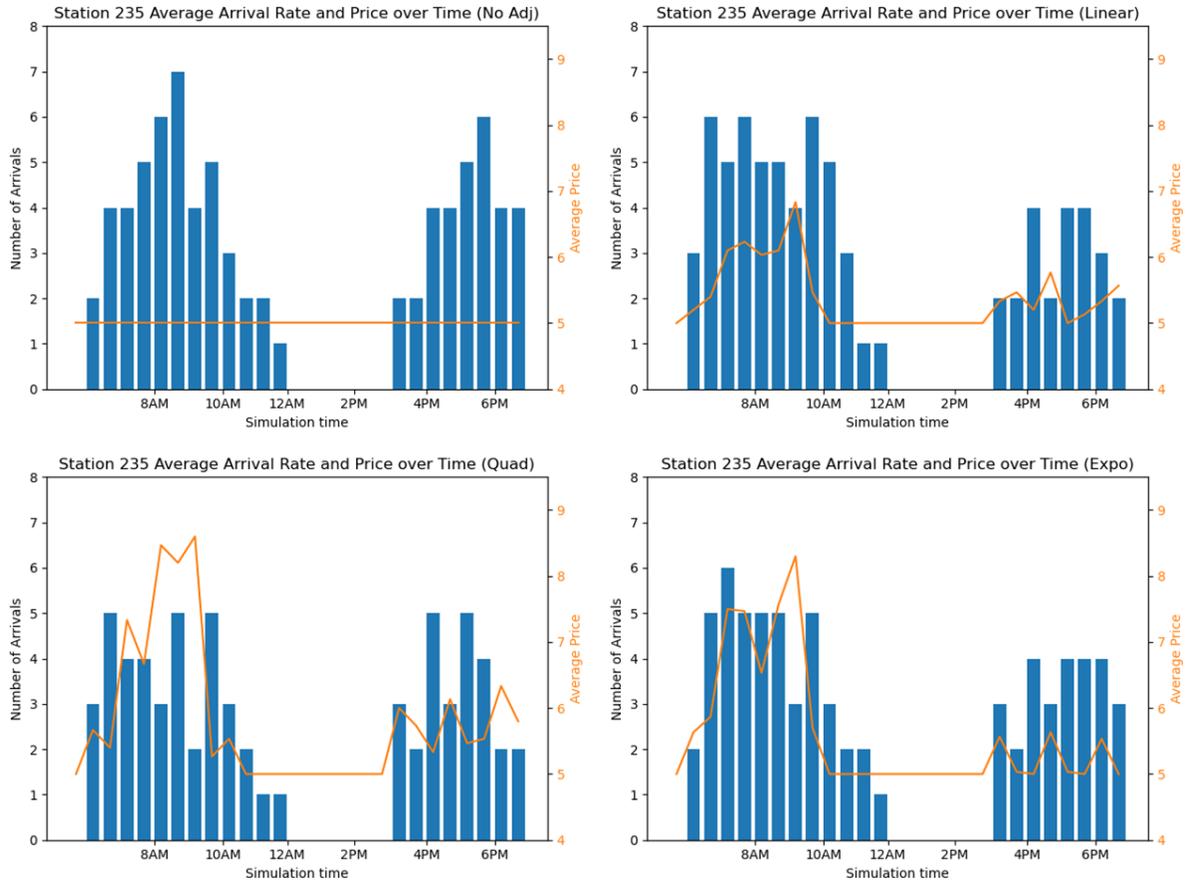

*Figure 10 Arrival rate and price*

Figure 11 displays the charging price, throughout the simulation period, for the five most congested charging stations across the four DDRPA schemes. Similar to the dynamics of queue length, prices increase during the peak period and remain at the base price during the off-peak periods of the day. It is worth mentioning that TOU pricing could be implemented to spread the overall demand for EV fast-charging across the day, in addition to DDRPA schemes that manage demand-supply imbalances in real-time. Overall, at the current demand level, the exponential adjustment scheme yields the most dramatic price changes, with prices nearing $20 per hour (3.5 times the base price) in some time intervals. The quadratic adjustment scheme generates a similar average price level as the exponential scheme but with less dramatic price fluctuations. The linear adjustment scheme results in significantly lower prices on average and significantly more stable prices throughout the analysis period.

For the no-adjustment pricing scheme, the price is constant at $5 throughout the entire analysis period. On average, for charging stations with congestion, the linear pricing adjustment scheme increases the hourly charging price by $1 ~ 3 dollars. The quadratic and exponential adjustment schemes increase charging price by $5 to $15 per hour over the no price adjustment case.

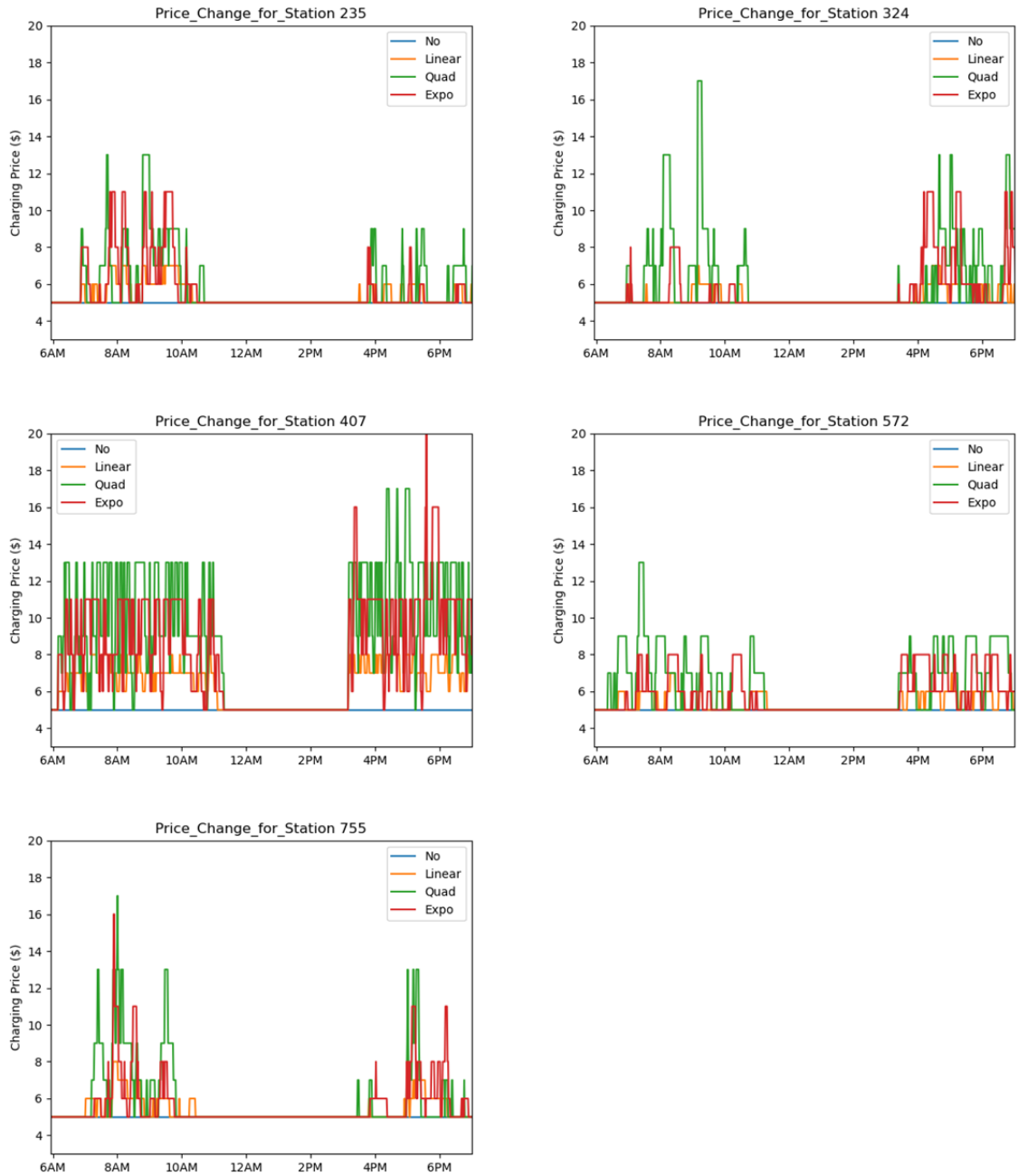

*Figure 11: Price Trend for Five Charging Stations*

## 5.4 Total Revenue

Table 5 displays the total system revenue results under the four DDRPA schemes; the table also displays the lost customer results to provide context. The results indicate that despite more lost customers under the quadratic price adjustment scheme, the scheme generates the highest revenue. Although the linear adjustment scheme does not generate as much revenue as the quadratic scheme, the linear scheme limits the number of lost customers. This indicates a clear trade-off between these two pricing schemes. On the contrary, the exponential adjustment scheme is strictly dominated by the linear adjustment scheme across the revenue and lost customer metrics.

**Table 5: System Revenue and Total Lost Customer Results**

|  | *No Adjustment* | *Linear* | *Quadratic* | *Exponential* |
|---|---|---|---|---|
| ***Total Revenue ($)*** | 6140 | 6371 | 6464 | 6350 |
| *Change relative to no-adjustment case* | N/A | 3.8% | 5.3% | 3.4% |
| ***Total Lost Customers*** | 511 | 547 | 583 | 574 |
| *Change relative to no-adjustment case* | N/A | 7.0% | 14.0% | 12.3% |

## 5.5 Customer Waiting Time

This subsection presents customer waiting time and customer total system time results and provides further details on the number of lost customers across the DDRPA schemes. Average waiting time and average total time in the system are both important metrics in the service industry. Table 6 summarizes the average waiting time, average total time in the system, and the number of lost customers and served customers across the four pricing schemes, where average wait time and average total time are calculated as follows:

$$Average\ Wait\ Time = \frac{\sum_{i \in C} Service\ Start\ Time_i - Station\ Arrival\ Time_i}{|C|} \quad (11)$$

$$Average\ Total\ Time = \frac{\sum_{i \in C} Service\ End\ Time_i - Station\ Arrival\ Time_i}{|C|} \quad (12)$$

where $C$ is the set of customers that receive service in the system.

**Table 6: Average Waiting Time and Total In-System Time Results**

|  | *No Adjustment* | *Linear* | *Quadratic* | *Exponential* |
|---|---|---|---|---|
| ***Avg. Waiting Time (min)*** | 6.38 | 5.94 | 4.74 | 4.91 |
| *Reduction over No-adjustment Case* | N/A | -7% | -26% | -23% |
| ***Avg. Total Time (min)*** | 26.48 | 26.09 | 24.97 | 25.05 |
| *Reduction over No-adjustment Case* | N/A | -1.5% | -5.7% | -5.4% |
| ***Served Customers*** | 3666 | 3630 | 3594 | 3603 |
| ***Lost Customers*** | 511 | 547 | 583 | 574 |
| ***Total Customers*** | | | 4177 | |

Table 6 shows that DDRPA schemes produce lower average wait times and lower average total system times than the no price adjustment case. The reductions in waiting time are around 7%, 26%, and 23% for the linear, quadratic and exponential schemes, respectively. Similarly, the reductions in total system time, relative to the no price adjustment case, are 1.5%, 5.7%, 5.4% for the linear, quadratic and exponential schemes, respectively. These results are not particularly surprising given the decrease in average queue lengths with the DDRPA schemes presented in Section 5.2—shorter queues equate to lower wait times.

Table 6 displays the total number of lost customers to illustrate the trade-offs with the DDRPA schemes. Unsurprisingly, the quadratic DDRPA scheme produces the lowest average wait times while serving the second lowest number of travelers. This pair of results is consistent as the fewer customers that are served, the shorter the average wait time for served customers should be.

## 5.6 Customer Disutility and Total Social Welfare

In the case study, the no-charging or balking option for users has a large constant disutility (-50). The preference to charge or not depends on the attributes of the charging station(s) in the user's choice set and the no-charging constant disutility. This subsection compares the average disutility for users (both served and lost users) in the regional EV fast-charging system. The following formula displays customer (dis-)utility:

$$U = \beta_1 \times Price + \beta_2 \times DetourDist + \beta_3 \times Waiting \quad (13)$$

In addition to comparing average disutility across the price adjustment schemes, this section also compares the total monetized utility (in dollar values) for all pricing schemes. The monetized utility of an individual customer is calculated as follows:

$$B = \begin{cases} -Price - \frac{\beta_2}{\beta_1} \times Detour - \frac{\beta_3}{\beta_1} \times Waiting, & \text{if a customer is served} \\ -\$18.5, & \text{if a customer leaves} \end{cases} \quad (14)$$

The monetized social welfare is the summation of charging system revenue and total customer monetized utility. This is the most holistic measure of system performance as it considers both the EV users service quality and the charging system revenue. Notably, every dollar of station revenue comes from a user paying for the station; hence, these two factors cancel out in the social welfare measure. The remaining factors that influence social welfare include wait times, detour distances, and the 'cost' of leaving the system. Hence, an effective pricing scheme should reduce wait time and detour costs more than it increases lost customer costs and detour costs.

Table 7 displays metrics for monetized average disutility of served users and all users, monetized total disutility of all users, total system revenue, and total social welfare. Comparing the three pricing cases with no-adjustment case, for the served customers who have only one or two station choices, the average disutility for charging decreases in all cases. For served customers who have three or more charging station alternatives, linear adjustment increases the disutility slightly, while quadratic and exponential adjustments both decrease the disutility. Therefore, for both quadratic and exponential price adjustment schemes, the average disutility for a served customer is lower than the no-adjustment case regardless of the customer's choice set. Overall, DDRPA schemes reduce served customer disutility from charging. The phenomenon results from the fact that waiting times and detour distances for served customers have decreased more than charging prices have increased, given the weights in the monetized utility function.

The grand average disutility and the total monetized customer utility rows in Table 7 indicate that for all customers—served and lost—the average disutility for the no-adjustment and linear cases are roughly the same level. Moreover, they indicate that the quadratic or exponential adjustment schemes reduce the average disutility of customers compared to the no-adjustment scheme. Hence, even though the quadratic and exponential adjustment schemes result in more lost travelers (each with a disutility of -$18.5) and higher costs for many served customers, the benefits of decreasing wait times outweigh these costs.

**Table 7: Customer Disutility and Social Welfare Results**

|  | Number of Charging Station Choices | Price Adjustment Schemes | | | |
|---|---|---|---|---|---|
|  |  | *No* | *Linear* | *Quadratic* | *Exponential* |
| *Monetized Average Disutility ($) for Served Customers* | 1 | 31.18 | 30.45 | 28.81 | 29.32 |
|  | 2 | 28.19 | 27.70 | 27.33 | 27.08 |
|  | ≥3 | 20.60 | 20.82 | 20.35 | 20.27 |
|  | Average Served | 24.91 | 24.69 | 23.90 | 23.98 |
| *Monetized Grand Average Disutility ($) \** | | 27.98 | 28.00 | 27.55 | 27.55 |
| *Monetized Total Disutility ($) \** | | -43,269 | -43,313 | -42,604 | -42,612 |
| *Total Revenue ($)* | | 6,140 | 6,371 | 6,464 | 6,350 |
| *Total Social Welfare ($)* | | -37,129 | -36,942 | -36,140 | -36,262 |

\* Includes served and lost customers

The last row of Table 7 is the most holistic metric for system evaluation as it combines a monetary measure of customer disutility and the revenue generated from the pricing schemes. According to the total social welfare results, each of the DDRPA schemes outperforms the no price adjustment case. The quadratic and exponential DDRPA schemes increase total social welfare by approximately 2.7% and 2.3%, respectively, compared to the no pricing case. Though not significant, the linear DDRPA scheme increases total social welfare by 0.5% compared to the no pricing case. Hence, these findings illustrate the potential benefits of using DDRPA schemes to improve a regional system of fast-charging stations, particularly the quadratic DDRPA scheme.

Table 8 summarizes the comparison between the three price adjustment schemes against the no-adjustment case. All DDRPA schemes achieve better system performance in reducing charging station queue length, waiting time and disutility, while at the same time increasing the total revenue and social welfare. From the system manager's perspective, using DDRPA schemes can significantly decrease queue lengths and generate higher revenues. From the customer perspective, DDRPA schemes can decrease wait time and average disutility.

Based on the social welfare measure, the quadratic scheme is the best DDRPA scheme. However, it is worth noting that the social welfare measure does not capture the distributional effects of DDRPA schemes. Hence, as pricing schemes tend to provide small benefits to many EV users (i.e., most users who remain in the system probably save one or two minutes in a queue) while providing large costs to a small number of EV users (i.e., some users do not get served and a small number likely pay significantly higher costs), the fact that the quadratic DDRPA produces significantly more lost customers than the linear DDRPA and the no pricing case is worth considering.

Table 8: Effects of Price-adjustment Schemes Compared to the No-adjustment Case (Base Scenario)

|  |  | No | Linear | Quadratic | Exponential |
|---|---|---|---|---|---|
| **Station** | Queue Length (Average) | 0.4 Person | -8% | -28% | -27% |
|  | Avg. Price (Congested Stations) | $5 | $7 | $10 | $8 |
|  | Price fluctuation | N/A | Stable | High | Medium |
|  | Total Revenue | $6,140 | +3.8% | +5.8% | +3.4% |
| **Customer** | Waiting Time | 6.4 min | -7% | -26% | -23% |
|  | Average Disutility (All) | 27.98 Unit | No change | -1.5% | -1.5% |
|  | Average Disutility (Served) | 24.91 Unit | -0.8% | -4% | -3.7% |
|  | Social Benefit | -37129 | +0.5% | +2.7% | +2.3% |
|  | Lost Customer | 511 | +3.8% | +5.3% | +3.4% |

*Note: Green Indicates advantages and red indicates disadvantages*

# 6 Scenario Analysis

This section applies the modeling framework to two other scenarios with different demand levels and compares the system performance metrics. The two new scenarios are a lower demand case and a higher demand case, compared to the base case presented in Section 4 and analyzed in Section 5. In the low demand scenario, EV users may not fully adapt to charging at public stations and still prefer charging at home or destination charging at a workplace or shopping center. In this case, this study assumes 50% of all EV users never use fast-charging stations, which reduces the market penetration of EV fast-charging trips from 1.4% to 0.7%. In the high demand scenario, EV sales are fast increasing, and public charging is fully adopted. In this case, the market penetration rate is assumed to double from 1.4% to 2.8%. All other simulation parameters remain the same. Table 9 and Table 10 display the system performance metrics for the low demand and high demand scenarios, respectively.

Table 9: Effects of Price-adjustment Schemes Compared to the No-adjustment Case (Low Demand Scenario)

**Market penetration 0.7%, all other parameters remains the same as Section 4**

|  |  | No | Linear | Quadratic | Exponential |
|---|---|---|---|---|---|
| **Station** | Queue Length (Average) | 0.3 Person | -5% | -30% | -21% |
|  | Avg. Price (Congested Stations) | $5 | $6 | $8 | $8 |
|  | Price fluctuation | N/A | Stable | Medium | Medium |
|  | Total Revenue | $3,199 | +1.9% | +1.9% | +3.1% |
| **Customer** | Waiting Time | 2.6 min | -7% | -35% | -18% |
|  | Average Disutility (All) | 23. Unit | No change | -0.3% | +0.2% |
|  | Average Disutility (Served) | 21.2 Unit | No change | -2.4% | -1.6% |
|  | Social Benefit | -14,578 | No change | +0.8% | +0.5% |
|  | Lost Customers | 131 | +1.5% | +6.6% | +6.6% |

*Note: Green Indicates advantages and red indicates disadvantages*

Table 10: Effects of Price-adjustment Schemes Compared to the No-adjustment Case (High Demand Scenario)

Market penetration 2.8%, all other parameters remains the same as Section 4

|  |  | No | Linear | Quadratic | Exponential |
|---|---|---|---|---|---|
| **Station** | Queue Length (Average) | 1.4 Person | -10% | -37% | -30% |
|  | Avg. Price (Congested Stations) | $5 | $8 | $12 | $12 |
|  | Price fluctuation | N/A | Medium | High | Very High |
|  | Total Revenue | $9,168 | +13.6% | +21.2% | +17.9% |
| **Customer** | Waiting Time | 15.1 min | -12% | -35.2% | -31.1% |
|  | Average Disutility (All) | 37.85 Unit | No change | -0.3% | -0.3% |
|  | Average Disutility (Served) | 31.55 Unit | -2% | -8% | -8% |
|  | Social Benefit | -10,5005 | +1.5% | +5.4% | +4.9% |
|  | Lost Customers | 2682 | +4.7% | +8.6% | +7.3% |

*Note: Green Indicates advantages and red indicates disadvantages*

The comparisons of the first columns (i.e., the no pricing case) in the three scenarios (Table 8/Table 9/Table 10) indicate that as the demand level increases, naturally congestion increases at the system level. The queue length, waiting time, average disutility, and lost customers all increase as demand level increases. Comparing Table 8 and Table 10, when demand doubles, the average queue length increases 3.5-fold and the waiting time more than doubles.

At relatively low demand levels (Table 8 and Table 9), the linear DDRPA scheme lowers the waiting time and maintains the number of lost customers at a low level. The price is also relatively stable under the linear DDRPA. On the other hand, quadratic and exponential DDRPA scheme achieve significant queue length and waiting time reductions, but the number of lost customers notably increase. Similar to the base demand case, the quadratic DDRPA results in the highest social welfare. However, it produces significantly more lost customers than the linear DDRPA, while only improving social welfare by 0.8% over the no pricing and linear DDRPA cases.

At high demand levels (Table 10), DDRPA schemes produce even larger benefits than at lower demand levels. The quadratic DDRPA scheme achieves the lowest queue length and waiting time and the highest total revenue and social benefit. The gap between the quadratic DDRPA and the linear DDRPA in terms of social welfare is quite significant—5.4% increase vs. 1.5%. Hence, for higher demand levels, more aggressive price adjustments appear to be more effective.

# 7 Conclusions

## 7.1 Summary

Motivated by the growing demand for EVs and EV fast-charging infrastructure, as well as the challenges preventing widespread access to EV fast-chargers, this study proposes a stochastic dynamic simulation modeling framework to support the real-time management and strategic planning of EV fast-charging stations. The modeling framework incorporates a discrete choice model for EV user station choice, queueing models to capture the dynamics and stochasticity of supply and demand at individual fast-charging stations, and dynamic demand-responsive price-adjustment (DDRPA) schemes to manage imbalances between supply and demand at individual stations.

Through a case study of a system of EV fast charging stations in Southern California, the study demonstrates that the proposed modeling framework can support (i) the real-time management of a system of EV fast-charging stations and (ii) the evaluation of the spatial allocation of fast-charging capacity in a region. The computational results illustrate that compared to the no DDRPA scheme case, the linear, quadratic, and exponential DDRPA schemes reduce queue lengths, user wait times, and customer disutility, while increasing revenues and overall social welfare. These findings hold under three different demand levels, with the benefits of DDRPA schemes increasing with the demand level. The social welfare measure incorporates total system revenue and customer disutility, as such, it is the most holistic measure of system performance. The quadratic DDRPA scheme outperforms the other DDRPA schemes and it increases social welfare compared to the no DDRPA scheme case by 0.8%, 2.7%, and 5.4% in the low, medium, and high demand scenarios, respectively.

In terms of the proposed modeling framework supporting evaluations of the spatial allocation of fast-charging capacity in a region, the case study clearly identifies two areas of the case study region that would benefit from additional capacity. In the first area, the northern part of the region, additional fast-charging stations would likely decrease lost customers, user wait times, and user detour distances and increase overall social welfare. In the second area, the eastern part of the region, while additional fast-charging stations may be beneficial, increasing the number of chargers at existing stations should produce the same improvements in system performance and social welfare.

In both the no DDRPA case and DDRPA scheme cases, the simulation results provide valuable insights into the EV users who did not use an EV fast-charging facility. In many cases, the EV user did not have any fast-charging stations in her choice set due to her EV's SOC and her willingness-to-detour. In other cases, the only feasible option was a single EV fast-charging station with either very long wait times (in the no DDRPA case) or very high prices (in the DDRPA cases).

## 7.2 Modeling Framework Applications

This section details the practical application of the modeling framework proposed in this study. There are various entities who may benefit from deploying the proposed modeling framework for decision support, under various circumstances. For example, it is possible (albeit unlikely) that EV manufacturers own and operate their own fast-charging stations and only provide charging compatibility to their own vehicles (e.g., Tesla, Ford, BMW, etc.). The modeling framework applies directly in this case. Another example is where a private company owns and operates a system of EV charging stations in a city, a suburb, or a heavily traveled corridor. In the case where this company has a monopoly or near monopoly in the region, the proposed model applies directly. In the case where the company is competing with other fast-charging providers, the proposed station choice behavioral model would probably need to be adjusted to incorporate this competition. Further research is needed to model competition between multiple providers of EV fast-charging within the same region. A final example includes a public sector owner-operator who manages a system of EV fast-charging stations or a governmental regulator who regulates a system of EV fast-charging stations. The proposed modeling framework would apply directly to their case as well.

One interesting modeling issue relates to the compatibility of EVs and fast chargers. Consider the case where a particular EV manufacturer manages a system of stations wherein the chargers are only compatible with their own EVs. In this case, to use the proposed modeling framework, the EV company would need to calibrate the input spatiotemporal demand such that only trips from their own EVs can use their system of charging stations, which they should be able to do relatively straightforwardly. Moreover, this company could use the proposed modeling framework to see how much revenue they are missing out on by not offering EV charging to all EVs. The EV manufacturer could also evaluate a tiered pricing system wherein their own EVs never pay the congestion surcharge but EVs from other manufacturers need to pay the congestion surcharge when queues exist. On a related note, a public sector regulator could use the proposed modeling framework to determine the social benefit of standardizing EV chargers to promote inter-operability.

Another valuable component of the proposed modeling framework in terms of applicability is that it is highly modular. There is a joint station choice and balking decision module; a multi-server queueing module; and station pricing module. Obviously, other researchers and practitioners can adjust the parameter values in each of these modules. Moreover, others can easily replace the multinomial logit model for station choice plus bulking decision with a nested logit, probit, or mixed logit model as well as a deterministic model. Similarly, they can replace the DDRPA schemes with significantly more advanced dynamic pricing policies that (i) are anticipatory rather than reactive, and (ii) consider all stations or a group of stations simultaneously rather than individually. An anticipatory pricing approach would require predictions of future demand at fast-charging stations; hence, an additional module would need to be incorporated into the modeling framework.

Lastly, the dynamic modeling framework is applicable to general types of fast-charging facilities, such as stations with Level 4 chargers. Model parameters as well as dynamic pricing policies would need to be adjusted to incorporate the shorter charging times associated with Level 4 chargers.

On the other hand, the proposed modeling framework is likely not directly applicable to the management of Level 2 chargers in workplace or shopping center parking lots because most users of these chargers are not charging en-route (or mid-trip) and they may leave their vehicles at chargers based on how long they are at work or shopping. While these Level 2 charging facilities would benefit from modeling and management, a simple multi-server queueing model may suffice.

### 7.3 Future Research

Given the novelty of modeling the dynamics of a system of EV fast charging stations, this study opens up a variety of interesting areas of future research. The first area, mentioned previously in the paper, is the integration of the dynamic modeling framework in this study with existing models in the literature that integrate the dynamics of EV charging infrastructure with electricity from the power grid. In this higher level of model integration, grid prices would most likely be treated as exogenous, and the EV charging station manager would determine when to purchase energy from the grid, when to store the energy at a station, and at what price to sell the energy to EVs considering the demand for EV fast-charging as well as the purchasing and storage price of energy.

A second area of future research involves a more direct integration of the dynamic modeling framework in this study with the deterministic optimization models of station sizing and siting. While sizing and siting models incorporate some level of supply-demand integration, they rarely incorporate stochasticity in EV user station choice, and they do not incorporate the temporal dynamics of supply and demand.

A minor shortcoming of the current study is that it uses the Manhattan distance metric to model detour distances. Incorporating a real-world road network for measuring detour distances would increase the realism of the proposed modeling framework.

Although not necessarily a shortcoming of the existing study, the current modeling framework assumes that EV users consider balking and station choice at the beginning of their trip and when they arrive at their chosen station. Future research may assume that EV users have access to a mobile phone application that updates expected station wait time and station prices regularly, thereby, allowing EV users to reconsider station choice when they are en-route to their originally chosen station.

A final area of future research involves modeling EV user tours/trip chains rather than individual trips. While the current study uses random number generation to determine the initial SOC of each EV user trip, a more detailed behavioral model that incorporates the entire sequence of activities (e.g. home, work, shopping, healthcare, eating out, etc.) and trips an EV user makes throughout the day, as well as a behavioral model of when and where these users plan to charge their EVs is possible. The downside of this activity-based modeling approach with EV user tours/trip chains is that it is significantly more computationally- and data-intensive than the current trip-based approach.

## Acknowledgements

The authors would like to thank the three anonymous reviewers for their valuable comments and suggestions on the initial manuscript and study. This feedback significantly improved the quality of the paper. Moreover, an earlier version of this paper was presented at the North American Regional Science Council (NARSC) 2020 virtual conference. The discussant for the paper was Dr. Kara Kockelman. We would like to thank Dr. Kockelman as well as Yantao Huang for their valuable feedback.

Pergamon, 38, pp. 44–55. doi: 10.1016/J.TRC.2013.11.001.

Edison Electric Institute (2019) 'Electric Vehicle Sales : Facts & Figures', pp. 1–4. Available at: https://www.eei.org/issuesandpolicy/electrictransportation/Documents/FINAL_EV_Sales_Update_Oct2019.pdf.

Egbue, O. and Long, S. (2012) 'Barriers to widespread adoption of electric vehicles: An analysis of consumer attitudes and perceptions', *Energy Policy*. Elsevier, 48(2012), pp. 717–729. doi: 10.1016/j.enpol.2012.06.009.

EVgo (2020) *EVgo Official Website*, *EVgo Official Website*. Available at: https://www.evgo.com/ (Accessed: 28 March 2020).

Farkas, C. and Prikler, L. (2017) 'Stochastic modelling of EV charging at charging stations', *Renewable Energy and Power Quality Journal*, pp. 1046–1051. doi: 10.24084/repqj10.574.

Flath, C. M. *et al.* (2014) 'Area Pricing Coordination Through Area Pricing', *Transportation Science*, 48(4), pp. 619–634.

Ge, Y. and MacKenzie, D. (2017) 'Dynamic discrete choice modeling of the charging choices of plug-in hybrid electric vehicle drivers', *Transportation Research Board 96th Annual Meeting*, (250), pp. 1–19.

Ge, Y., MacKenzie, D. and Keith, D. R. (2018) 'Gas anxiety and the charging choices of plug-in hybrid electric vehicle drivers', *Transportation Research Part D: Transport and Environment*. Elsevier, 64(June 2017), pp. 111–121. doi: 10.1016/j.trd.2017.08.021.

Ghamami, M., Zockaie, A. and Nie, Y. (Marco) (2016) 'A general corridor model for designing plug-in electric vehicle charging infrastructure to support intercity travel', *Transportation Research Part C: Emerging Technologies*. Pergamon, 68, pp. 389–402. doi: 10.1016/J.TRC.2016.04.016.

Gross, D. *et al.* (2008) *Fundamentals of Queueing Theory*. 4th edn. John Wiley & Sons, Inc. doi: 10.1002/9781118625651.

Guo, Z., Deride, J. and Fan, Y. (2016) 'Infrastructure planning for fast-charging stations in a competitive market', *Transportation Research Part C: Emerging Technologies*. Elsevier Ltd, 68, pp. 215–227. doi: 10.1016/j.trc.2016.04.010.

He, F. *et al.* (2013) 'Optimal deployment of public charging stations for plug-in hybrid electric vehicles', *Transportation Research Part B: Methodological*, 47(2013), pp. 87–101. doi: 10.1016/j.trb.2012.09.007.

He, F., Yin, Y. and Zhou, J. (2015) 'Deploying public charging stations for electric vehicles on urban road networks', *Transportation Research Part C: Emerging Technologies*. Elsevier Ltd, 60, pp. 227–240. doi: 10.1016/j.trc.2015.08.018.

Hu, L., Dong, J. and Lin, Z. (2019) 'Modeling charging behavior of battery electric vehicle drivers: A cumulative prospect theory based approach', *Transportation Research Part C: Emerging Technologies*. Elsevier, 102(August 2018), pp. 474–489. doi: 10.1016/j.trc.2019.03.027.

Huang, Y. and Kockelman, K. M. (2020) 'Electric vehicle charging station locations: Elastic demand, station congestion, and network equilibrium', *Transportation Research Part D: Transport and Environment*. Elsevier, 78(November 2019), p. 102179. doi: 10.1016/j.trd.2019.11.008.

Iyer, V. M. *et al.* (2018) 'Extreme fast-charging station architecture for electric vehicles with partial power processing', *Conference Proceedings - IEEE Applied Power Electronics Conference and Exposition - APEC*. IEEE, 2018-March, pp. 659–665. doi: 10.1109/APEC.2018.8341082.

Jabeen, F. *et al.* (2013) 'Electric vehicle battery charging behaviour : findings from a driver survey', *Australasian Transport Research Forum Proceedings 2013*, pp. 1–15.

Jain, R. and Smith, J. M. (2008) 'Modeling Vehicular Traffic Flow using M/G/C/C State Dependent Queueing Models', *Transportation Science*, 31(4), pp. 324–336. doi: 10.1287/trsc.31.4.324.

Kang, J. E. and Recker, W. W. (2009) 'An activity-based assessment of the potential impacts of plug-in hybrid electric vehicles on energy and emissions using 1-day travel data', *Transportation Research Part D: Transport and Environment*. Elsevier Ltd, 14(8), pp. 541–556. doi: 10.1016/j.trd.2009.07.012.

Khan, W., Ahmad, F. and Alam, M. S. (2019) 'Fast EV charging station integration with grid ensuring optimal
35